\documentclass[12pt,a4paper]{article}

\usepackage{amsmath,amsfonts,latexsym,amssymb,amscd}
\usepackage{pslatex}
\usepackage[latin1]{inputenc}
\usepackage[T1]{fontenc}
\usepackage{pspicture}
\usepackage{verbatim,amsthm,curves,graphics}
\usepackage{mathrsfs}

\def\ben{\begin{equation}}

\def\een{\end{equation}}

\def\bena{\begin{eqnarray}}

\def\eena{\end{eqnarray}}

\def\f(#1/#2){\frac{#1}{#2}}

\def\Frac(#1/#2){\left(\frac{#1}{#2}\right)}

\def\chris(#1-#2-#3){{\mit \Gamma}^{#1}{}_{{#2}{#3}} }

\def\tilchris(#1-#2-#3){\tilde{{\mit \Gamma}}^{#1}{}_{{#2}{#3}}}

\def\hatchris(#1-#2-#3){\hat{{\mit \Gamma}}^{#1}{}_{{#2}{#3}}}

\newfont{\schwell}{schwell scaled 1101}

\newcommand{\non}{\nonumber}

\theoremstyle{definition}

\newtheorem{thm}{Theorem}

\newtheorem{lemma}{Lemma}

\renewcommand{\epsilon}{\varepsilon}
\newcommand{\myid}{{\bf 1}}
\newcommand{\mr}{{\mathbb R}}

\newcommand{\mc}{{\mathbb C}}

\renewcommand{\i}{{\rm i}}
\newcommand{\supp}{\operatorname{supp}}

\newcommand{\I}{{\mathcal I}}

\newcommand{\T}{{\bf T}}

\newcommand{\M}{{\bf M}}
\newcommand{\C}{{\mathcal C}}
\newcommand{\sd}{\operatorname*{sd}}

\renewcommand{\O}{{\mathcal O}}
\newcommand{\WF}{\operatorname{WF}}
\newcommand{\bv}{\operatorname*{B.V.}}

\renewcommand{\d}{{\rm dim}}

\renewcommand{\H}{{\mathcal H}}
\newcommand{\x}{{\vec x}}

\newcommand{\A}{{\mathcal A}}
\renewcommand{\S}{{\mathcal S}}
\begin{document}

\title{Axiomatic quantum field theory in curved spacetime}

\author{Stefan Hollands$^a$\thanks{HollandsS@Cardiff.ac.uk} \,, Robert
  M. Wald$^b$\thanks{rmwa@midway.uchicago.edu} \\
  \\
  $^a$ {\em School of Mathematics, Cardiff University, UK} \\
  \\
  $^b$ {\em Enrico Fermi Institute and Department of Physics} \\
{\em University of Chicago, Chicago, IL, USA}
  \\
  \\
  \\}

\maketitle

\abstract{
The usual formulations of quantum field theory in Minkowski spacetime
make crucial use of features---such as Poincare invariance and the
existence of a preferred vacuum state---that are very special to
Minkowski spacetime. In order to generalize the formulation of quantum
field theory to arbitrary globally hyperbolic curved spacetimes, it is
essential that the theory be formulated in an entirely local and
covariant manner, without assuming the presence of a preferred
state. We propose a new framework for quantum field theory, in which
the existence of an Operator Product Expansion (OPE) is elevated to a
fundamental status, and, in essence, all of the properties of the
quantum field theory are determined by its OPE. We provide general
axioms for the OPE coefficients of a quantum field theory. These
include a local and covariance assumption (implying that the quantum
field theory is locally and covariantly constructed from the spacetime
metric), a microlocal spectrum condition, an
"associativity" condition, and the requirement that
the coefficient of the identity in the OPE of the product of a field
with its adjoint have positive scaling degree. We prove curved spacetime
versions of the spin-statistics theorem and the PCT theorem. Some
potentially significant further
implications of our new viewpoint on quantum field theory are discussed.  }

\section{Introduction}

The Wightman axioms \cite{sw} of quantum field theory in Minkowski
spacetime are generally believed to express the fundamental properties
that quantum fields possess. In essence, these axioms require that the
following key properties hold: (1) The states of the theory are unit
rays in a Hilbert space, ${\mathcal H}$, that carries a unitary
representation of the Poincare group. (2) The 4-momentum (defined by
the action of the Poincare group on the Hilbert space) is positive,
i.e., its spectrum is contained within the closed future light cone
(``spectrum condition''). (3) There exists a unique, Poincare
invariant state (``the vacuum''). (4) The quantum fields are
operator-valued distributions defined on a dense domain
${\mathcal D} \subset {\mathcal H}$ that is both
Poincare invariant and invariant under the action of the fields and
their adjoints. (5) The fields transform in a covariant manner under
the action of Poincare transformations. (6) At spacelike separations,
quantum fields either commute or anticommute.

During the past 40 years, considerable progress has been made in
understanding both the physical and mathematical properties of quantum
fields in curved spacetime. Although gravity itself is treated
classically, this theory incorporates some key aspects of general
relativity and thereby should provide a more fundamental base for
quantum field theory. Much of the progress has occurred in the
analysis of free (i.e., non-self-interacting) fields, but in the past
decade, major progress also has been made in the perturbative analysis
of interacting quantum fields. Significant insights have thereby been
obtained into the nature of quantum field phenomena in strong
gravitational fields. In addition, some important insights have been
obtained into the nature of quantum field theory itself. One of the
key insights is that---apart from stationary spacetimes or spacetimes
with other very special properties---there is no unique, natural notion of a
``vacuum state'' or of ``particles''. Indeed, unless the spacetime is
asymptotically stationary at early or late times, there will not, in
general, even be an asymptotic notion of particle states. Consequently,
it is essential that quantum field theory in curved spacetime be formulated in
terms of the local field observables as opposed, e.g., to S-matrices.

Since quantum field theory in curved spacetime should be much closer
to a true theory of nature than quantum field theory in Minkowski
spacetime, it is of interest to attempt to abstract the fundamental
features of quantum field theory in curved spacetime in a manner
similar to the way the Wightman axioms abstract what are generally
believed to be the fundamental features of quantum field theory in
Minkowski spacetime. The Wightman axioms are entirely compatible with
the focus on local field observables, as needed for a formulation of
quantum field theory in curved spacetime. However, most of the
properties of quantum fields stated in the Wightman axioms are very
special to Minkowski spacetime and cannot be generalized straightforwardly to
curved spacetime. Specifically, a curved spacetime cannot possess
Poincare symmetry---indeed a generic curved spacetime will not possess
any symmetries at all---so one certainly cannot require ``Poincare
invariance/covariance'' or invariance under any other type of
spacetime symmetry. Thus, no direct analog of properties (3) and (5)
can be imposed in curved spacetime, and the key aspects of properties
(1) and (2) (as well as an important aspect of (4)) also do not make
sense.

In fact, the situation with regard to importing properties (1), (2),
and (4) to curved spacetime is even worse than would be suggested by
merely the absence of symmetries: There exist unitarily inequivalent
Hilbert space constructions of free quantum fields in spacetimes with
a noncompact Cauchy surface and (in the absence of symmetries of the
spacetime) none appears ``preferred''. Thus, it is not appropriate
even to assume, as in (1), that states are unit rays in a {\it single}
Hilbert space, nor is it appropriate to assume, as in (4), that the
(smeared) quantum fields are operators on this unique Hilbert
space. With regard to (2), although energy and momentum in curved
spacetime cannot be defined via the action of a symmetry group, the
stress-energy tensor of a quantum field in curved spacetime should be well
defined as a distributional observable on spacetime, so one might hope that it
might be possible to, say, integrate the (smeared) energy density of a quantum
field over a Cauchy surface and replace the Minkowski spacetime
spectrum condition by the condition that the total energy of the
quantum field in any state always is non-negative. However, this is
not a natural thing to do, since the ``total energy'' defined in this
way is highly slice/smearing dependent, and it is well known in classical
general relativity that in asymptotically flat spacetimes, the
integrated energy density of matter may bear little relationship to
the true total mass-energy.  Furthermore, it is well known that the
energy density of a quantum field (in flat or curved spacetime) can be
negative, and, in some simple examples involving free fields in curved
spacetime, the integrated energy density is found to be
negative. Consequently, there is no analog of property (2) in curved
spacetime that can be formulated in terms of the ``total
energy-momentum'' of the quantum field. Thus, of all of the properties
of quantum fields in Minkowski spacetime stated in the Wightman
axioms, only property (6) has a straightforward generalization to
curved spacetimes!

Nevertheless, it has been understood for quite some time that the
difficulties in the formulation of quantum field theory in curved
spactime that arise from the existence of unitarily inequivalent
Hilbert space constructions of the theory can be overcome by simply
formulating the theory via the algebraic framework~\cite{haag}. Instead of
starting from the postulate that the states of the theory comprise a
Hilbert space and that the (smeared) quantum fields are operators on
this Hilbert space, one starts with the assumption that the (smeared)
quantum fields (together with the identity element $\myid$)
generate a *-algebra, $\A$. States are then
simply expectation functionals
$\langle \, . \, \rangle_\omega:\A \rightarrow \mc$ on the algebra,
i.e., linear functionals that
are positive in the sense that $\langle A^* A \rangle_\omega \geq 0$
for all $A \in \A$. The GNS construction then assures us that given a
state, $\omega$, one can find a Hilbert space ${\mathcal H}$ that
carries a representation, $\pi$, of the *-algebra $\A$, such
that there exists a vector $|\Psi \rangle \in {\mathcal H}$ for which
$\langle A \rangle_\omega = \langle \Psi|\pi(A)|\Psi \rangle$ for all $A \in
\A$. All of the operators, $\pi(A)$, are automatically
defined on a common dense invariant domain, ${\mathcal D} \subset
{\mathcal H}$, and each vector
$|\Psi \rangle\in {\mathcal D}$ defines a state via
$\langle A \rangle_\Psi = \langle \Psi|\pi(A)|\Psi \rangle$. Thus, by simply
adopting the algebraic viewpoint, we effectively incorporate into
quantum field theory in curved spacetime the portions of the content of
properties (1) and (4) above that do not refer to Poincare symmetry.

It is often said that in special relativity one has invariance under
``special coordinate transformations'' (i.e., Poincare
transformations), whereas in general relativity, one has invariance
under ``general coordinate transformations'' (i.e., all
diffeomorphisms). Thus, one might be tempted to think that the
Minkowski spacetime requirements of invariance/covariance under
Poincare transformations could be generalized to curved spacetime by
requiring a similar ``invariance/covariance under arbitrary
diffeomorphisms''. However, such thoughts are based upon a
misunderstanding of the true meaning of ``special covariance'' and
``general covariance''. By explicitly incorporating the flat spacetime
metric, $\eta_{ab}$, into the formulation of special relativity, it can
easily be seen that special relativity can be formulated in as
``generally covariant'' a manner as general relativity. However, the
act of formulating special relativity in a generally covariant manner
does not provide one with any additional symmetries or other useful
conditions on physical theories in flat spacetime. The point is that
in special relativity, Poincare transformations are {\it symmetries}
of the spacetime structure, and we impose a nontrivial requirement on
a physical theory when we demand that its formulation respect these
symmetries. However, a generic curved spacetime will not possess any
symmetries at all, so no corresponding conditions on a physical theory can be
imposed. The demand that a theory be ``generally covariant'' (i.e.,
that its formulation is invariant under arbitrary diffeomorphisms) can
always be achieved by explicitly incorporating any ``background
structure'' into the formulation of the theory. If one
considers a fixed, curved spacetime without symmetries, no useful
conditions can be imposed upon a quantum field theory by attempting to
require some sort of ``invariance'' of the theory under
diffeomorphisms.

However, there {\it is} a meaningful notion of ``general covariance''
that can be very usefully and powerfully applied to quantum field
theory in curved spacetime. The basic idea behind this notion is that the
only ``background structure'' that should occur in the theory is the
spacetime manifold and metric modulo diffeomorphisms, together with
the time and space orientations and (if spinors are present in the
theory) spin structure. The quantum fields should be ``covariant'' in
that they should be constructed from only this background structure. Indeed,
since the smeared quantum fields are associated with local regions of
spacetime (namely, the support of the test function used for the
smearing), it seems natural to demand that the quantum fields be {\it
locally} constructed from the background structure in the sense that
the quantum fields in any neighborhood ${\mathcal O}$ be covariantly
constructed from the background structure within ${\mathcal O}$. This idea
may be formulated in a precise manner as follows~\cite{hw1,hw2,bfv}.

First, in order to assure a well defined dynamics and in order to
avoid causal pathologies, we restrict consideration to globally
hyperbolic spacetimes $(M,g_{ab})$. (We consider theories in arbitrary
spacetime dimension $D \equiv{\rm dim} \, M \geq 2$.) If spinors are
present in the theory, we also demand that $M$ admit a spin
structure. It is essential that the quantum field theory
be defined on {\it all} $D$-dimensional globally hyperbolic
spacetimes admitting a spin structure, since in essence, we can only
tell whether the quantum field is ``locally and covariantly
constructed out of the metric'' if we can see how the theory changes
when we change the metric in an arbitrary way.
The ``background structure'', $\M$, of the
theory is taken to consist of the manifold $M$, the metric
$g_{ab}$, the spacetime orientation---which may be
represented by a nowhere vanishing $D$-form, $e_{a_1 \dots a_D}$
on $M$---and a time
orientation---which may be represented e.g. by the equivalence class
of a time function $T: M
\to \mr$---i.e., we have
\ben
\M = (M,g,T,e) \, .
\een
(If spinors
are present in the theory, and $M$ admits more than one spin
structure, then the choice of spin-structure over $M$ also should be
understood to be included $\M$.) For each choice of $\M$, we assume
that there is specified a *-algebra $\A(\M)$ that is generated by a
countable list of quantum fields $\phi^{(i)}$ and their ``adjoints''
$\phi^{(i)*}$. These fields may be of arbitrary tensorial or spinorial
type, and they are smeared with arbitrarily chosen smooth, compact
support fields of dual tensorial or spinorial type. In order to
determine if the quantum field theory and quantum fields $\phi^{(i)}$
are ``locally and covariantly constructed out of the background
structure $\M$'', we consider the following situation: Let $(M,g)$ and
$(M',g')$ be two globally hyberbolic spacetimes that have the property
that there exists a one-to-one (but not necessarily onto) map $\rho: M
\to M'$ that preserves all of the background structure. In other
words, $\rho$ is an isometric imbedding that is orientation and time
orientation preserving (and, if spinors are present, the choices of
spin structure on $M$ and $M'$ correspond under $\rho$). We further
assume that $\rho$ is causality preserving in the sense that if $x_1,
x_2 \in M$ cannot be connected by a causal curve in $M$, then
$\rho(x_1)$ and $\rho(x_2)$ cannot be connected by a causal curve in
$M'$. We say that the quantum field theory is {\it locally and
covariantly constructed from $\M$} (or, for short, that the theory
is {\it local and covariant}) if (i) for every such $\M$, $\M'$, and
$\rho$ we have a corresponding *-isomorphism $\chi_\rho$ between
$\A(\M)$ and the subalgebra of $\A(\M')$ generated by the quantum
fields $\phi^{(i)}$ and $\phi^{(i)*}$ smeared with test fields with
support in $\rho[M]$ and (ii) if $\rho'$ is a similar background
structure and causality preserving map taking $\M'$ to $\M''$, then
$\chi_{\rho' \circ \rho} = \chi_{\rho'} \circ \chi_\rho$. We further
say that the quantum field $\phi^{(i)}$ is {\it locally and
  covariantly constructed from $\M$} (or, for short, that $\phi^{(i)}$
is {\it local and covariant}) if for every such $\M$, $\M'$, and
$\rho$, we have \ben \chi_\rho \Big(\phi^{(i)} (f) \Big) = \phi^{(i)}
(\rho_* (f)) \quad ,
\label{lcfield}
\een
where $\rho_*(f)$ denotes the natural push-forward action of $\rho$ on the
tensor/spinor field $f$ on $M$.

Note that in contrast to the notion of Poincare
invariance/covariance---which applies to quantum field theory on a
single spacetime (namely, Minkowski spacetime)---the notion that a
quantum field theory or quantum field is local and covariant is a
condition that applies to the formulation of quantum field theory on
{\it different} spacetimes. Nevertheless, the close relationship
between these notions can be seen as follows: Suppose that we have a
local and covariant quantum field theory, with local and covariant
quantum fields $\phi^{(i)}$. Let $\M$ and $\M'$ both be the background
structure of Minkowski spacetime, and let $\rho$ be a proper Poincare
transformation. Then $\rho$ preserves all of the background structure,
so for each proper Poincare transformation, we obtain a *-isomorphism
$\chi_\rho: \A \to \A$, where $\A$ here
denotes the quantum field algebra for Minkowski
spacetime. Furthermore, if $\rho$ and $\rho'$ are proper Poincare
transformations, we have $\chi_{\rho \circ \rho'} =
\chi_{\rho} \circ \chi_{\rho'}$.
Thus, every local and covariant
quantum field theory in curved spacetime gives rise to a Poincare
invariant theory in this sense when restricted to Minkowski spacetime.
Furthermore, if $\phi^{(i)}$ is a local and covariant quantum field,
then in Minkowski spacetime it transforms covariantly via
eq.(\ref{lcfield}) under proper Poincare transformations.

Note also that, more generally, in any curved
spacetime with symmetries, a local and covariant quantum field theory
will be similarly invariant under these symmetries, and a local and covariant
quantum field will transform covariantly under these symmetries. But
even for spacetimes without any symmetries, the requirement that the
quantum field theory and quantum fields be local and covariant imposes
a very powerful restriction akin to requiring Poincare invariance/covariance
in Minkowski spacetime.

From these considerations, it can be seen that if we adopt the above
algebraic framework for quantum field theory in curved spacetime and
if we additionally demand that the quantum field theory and the
quantum fields $\phi^{(i)}$ be local and covariant, then we obtain
satisfactory generalizations to curved spacetime of
properties (1), (4), and (5) of the Wightman axioms in Minkowski
spacetime. Since we already noted that (6) has a trivial
generalization to curved spacetime, only properties (2) and (3)
remain to be generalized to curved spacetime.

We have already noted above that there is no analog of property (2) in
curved spacetime that can be formulated in terms of the ``total
energy-momentum'' of the quantum field. However, it is possible to
reformulate the spectrum condition in Minkowski spacetime in terms of
purely local properties of the quantum fields.  Specifically, the
``positive frequency'' (and, thereby, positive energy) properties of
states are characterized by the short-distance singularity structure
of the $n$-point functions of the quantum fields, as described by their
wavefront set. One thereby obtains a {\it microlocal spectrum
condition}~\cite{radzikowski,bfk,bf} that is formulated purely in terms of the local in
spacetime properties of the quantum fields. This microlocal spectrum
condition has a natural generalization to curved spacetime (see section~\ref{section2}
below), thus providing the desired generalization of property (2) to
curved spacetime.

Consequently, only property (3) remains to be generalized. In
Minkowski spacetime, the existence of a unique, Poincare invariant
state has very powerful consequences, so it is clear that a key portion
of the content of quantum field theory in Minkowski spacetime would be
missing if we failed to impose an analogous condition in curved
spacetime. However, as already mentioned above, one of the clear
lessons of the study of free quantum fields in curved spacetime is
that, in a general curved spacetime, there does not exist a unique,
``preferred'' vacuum or other state. Furthermore, even if a
prescription for finding a unique ``preferred state'' on each spacetime could
be found, since generic curved spacetimes do not have any symmetries
and states on different spacetimes cannot be meaningfully compared,
there would appear to be no sensible ``invariance'' properties that
such a preferred state could have. We do not believe that property (3)
can be generalized to curved spacetime by a condition that postulates
the existence of a preferred state with special properties.

In addition, we question the fundamental status of demanding the
existence of a state that is invariant under the symmetries of the
spacetime. For example, it is well known that the free massless
Klein-Gordon field in two-dimensional Minkowski spacetime does not
admit a Poincare invariant state. However, there is absolutely nothing
wrong with the quantum field algebra of this field; the quantum field
theory is ``Poincare invariant'' and the quantum field is ``Poincare
covariant'' in the sense described above. Furthermore, there is no
shortage of physically acceptable (``Hadamard'') states. Thus, the
only thing unusual about this quantum field theory is that it happens
not to admit a Poincare invariant state. We do not feel that this is
an appropriate reason to exclude the free massless Klein-Gordon field
in two-dimensional Minkowski from being considered to be a legitimate
quantum field theory. Similar remarks apply to the free Klein-Gordon
field of negative $m^2$ in Minkowski spacetime of all dimensions. The
classical and quantum dynamics of this field are entirely well posed
and causal, although they are unstable in the sense of admitting
solutions/states where the field grows exponentially with time. This
instability provides legitimate grounds for arguing that the free
Klein-Gordon field of negative $m^2$ does not occur in nature, but we
do not feel that the absence of a Poincare invariant state constitutes
legitimate grounds for rejecting this theory as a quantum field theory;
see section 6 below for further discussion.

For the above reasons, we seek a replacement of property (3) that does
not require the existence of states of a special type.  The main
purpose of this paper is to propose that the appropriate replacement
of property (3) for quantum field theory in curved spacetime is to
{\em postulate} the existence of a suitable {\it operator product expansion}~\cite{wilson,zimmermann,schroer}
of the quantum fields. The type of operator product expansion that we
shall postulate is known to hold in free field theory and to hold
order by order in peturbation theory on any
Lorentzian curved spacetime~\cite{hollands2006}. We thus propose to elevate this
operator product expansion to the status of a fundamental property of
quantum fields\footnote{
Various axiomatic approaches to conformal field theories based on the
operator product expansion have been proposed previously, see e.g.
the one based upon the notion of "vertex operator algebras" given
in~\cite{vertex1,vertex2,vertex3}. However, in contrast to our approach,
these approaches incorporate in an essential way the conformal symmetry
of the underlying space. In some approaches to quantum field theory
on Minkowski space, the OPE is not postulated, but instead
derived~\cite{fre,Bostelmann1,Bostelmann2}.
}. Although the assumption of the existence of an
operator product expansion in quantum field theory in curved spacetime
is remarkably different in nature from the assumption of the existence
of a Poincare invariant state in quantum field theory in Minkowski
spacetime, we will show in sections~\ref{section4} and~\ref{section5}
below that it can do ``much of
the same work'' as the latter assumption. It is shown in~\cite{hollands08}
how to exploit consistency conditions on the OPE
in a framework closely related to that presented here. In
particular, it is shown how
perturbations of a quantum field theory can be
characterized and calculated via consistency conditions
arising from the OPE.

We have an additional motivation for proposing to elevate the operator
product expansion to the status of a fundamental property of quantum
fields.  For free quantum fields in curved spacetime, an entirely
satisfactory *-algebra, $\A_0$, of observables has been
constructed~\cite{bfk,bf,hw1}, which includes all Wick powers and time-ordered
products. However, the elements of $\A_0$ correspond to
unbounded operators, and there does not seem to be any natural algebra
of bounded elements (with, e.g., a $C^*$-structure) corresponding to
$\A_0$. Furthermore, $\A_0$ does not appear to
have any natural topology (apart from a topology that can be defined
{\it a postiori} by using the allowed states as semi-norms).
Fortunately, a topology is not actually needed to define $\A_0$ because the relations that hold in $\A_0$ can be
expressed in terms of finite sums of finite products of
fields. However, it is inconceivable that the relations that define an
interacting field algebra will all be expressible in terms of finite
sums of finite products of the fields. Thus, without a natural
topology and without finitely expressible relations, it is far from
clear as to how an interacting field algebra might be defined.  We
claim that an operator-product expansion effectively provides the needed
``relations'' between the quantum fields, and we will propose in this paper
that these relations are
sufficient to define a quantum field theory. In other
words, we believe that an interacting quantum field theory is, in essence,
{\it defined} via its operator-product expansion. From this perspective, it
seems natural to view the operator product-expansion as a fundamental
aspect of the quantum field theory.

In the next section, we will describe our framework for quantum field
theory in curved spacetime. In particular, we will provide a precise
statement of the what we mean by an operator-product expansion and the
properties that we will assume that it possesses. In
section~\ref{section3}, we will state our axioms for quantum field
theory in curved spacetime and explain how it is constructed from the
operator-product expansion. Finally, in sections~\ref{section4}
and~\ref{section5} we will show that our axioms have much of the same
power as the Wightman axioms by establishing ``normal''
(anti-)commutation relations and proving curved spacetime versions of
the spin-statistics theorem and PCT theorem. Some further implications
of our new perspective on quantum field theory are discussed in
section 6.

\section{General framework for the formulation of QFT}\label{section2}

We will now explain in much more detail our proposed framework
for defining quantum field theory.
We fix a dimension $D \ge 2$ of the spacetime and consider all $D$-dimensional
globally hyperbolic spacetimes $(M,g_{ab})$. As explained
in the previous section, we will
assume that each spacetime is equipped with
an orientation, specified by
a nowhere vanishing $D$-form $e_{a_1 \dots a_D}$
on $M$, and a time orientation, specified by (the equivalence class of)
a globally defined time function
$T: M \to \mr$. The set of background data specified this way will be denoted
\ben
\M = (M,g,T,e) \, .
\een
In certain cases, more background structure may be prescribed, such as a
choice of bundles in which the quantum fields live. For
example, if spinors are present and if $M$ admits more than one spin
structure, then a choice of spin-structure
over $M$ is assumed to be given as part of the background
structure\footnote{It is convenient to think of the background structure as
a category~\cite{bfv}, whose objects are the tuples $\M$.
Morphisms in the
category of tuples $\M$ are isometric, causality, orientation, and other
background structure preserving embeddings
\ben
\rho: \M \to \M' \, .
\een
Thus, $\rho$ is a diffeomorphism $M \to M'$ such that $g = \rho^* g'$,
such that $\rho^* T'$ represents the same time-orientation as $T$, such
that $\rho^* e'$ represents the same orientation as $e$,
and such that the causal relations in $(M,g)$ inherited from $(M',g')$
coincide with the original ones.  Furthermore, if $\M$ also includes the
choice of a spin structure, then $\rho$ must also
preserve the spin structures.},
and is understood to also be part of $\M$. It should be emphasized that
two spacetimes with the
same manifold and metric, but, e.g.,
with different time-orientations define distinct
background structures.
Below, we will
consider quantum field theories associated with background structures, and
we stress that, at this stage, the quantum field theories associated with
different background structures (e.g., ones that merely differ in the
choice of, say, time-orientation)
need not have any relation whatsoever.

\medskip
The quantum fields present in a given theory will be assumed to
correspond to sections of vector bundles over $M$. We will denote the
various quantum fields by $\phi^{(i)}$, with $i \in \I$ where $\I$
is a suitable indexing set, and we will write $V(i)$ for the vector
bundle over $M$, of which $\phi^{(i)}$ corresponds to a section. It
should be emphasized that $i \in \I$ labels all of the quantum
fields present in the theory, not just the ``fundamental'' ones. Thus,
even if we were considering the theory of a single scalar field
$\varphi$, there will be infinitely many composite fields of various
tensorial types corresponding to all monomials in $\varphi$ and its
derivatives, each of which would be labeled by a different index
$i$. It will be convenient to also include a field denoted
$\phi^{(\myid)}$ in the list of quantum fields, which will play the role
of the identity element, $\myid$, in the quantum field algebra.

We assume
that each field $\phi^{(i)}$ has been assigned a Bose/Fermi parity
$F(i) = 0,1$ modulo two.

We further assume that there is an operation
\ben \star: I \to I \quad i \mapsto i^\star \, ,
\een
having the property that $V(i^\star)=\overline V(i)$, where
for any vector space $E$, the vector space $\overline E$ consists of
all anti-linear maps $E^{\rm v} \to \mc$, with $E^{\rm v}$ denoting
the dual space of $E$.  In particular, if $i$ is associated with,
say the vector bundle $V(i)$ of spinors with $P$ primed and $U$
unprimed indices, then $V(i^\star)$ is the bundle of spinors
with $U$ primed and $P$ unprimed spinor indices. We demand that the
star operation squares to the identity\footnote{This operation gives
$\I$ the structure of an involutive category.}, $i^{\star\star}=i$.
We also require that $\phi^{(\myid^*)} = \phi^{(\myid)}$, i.e. that
$\myid^* = \myid$.

As in many other approaches to quantum field theory, we will use the
smeared fields $\phi^{(i)}(f)$---with $i \in \I$, and $f$ a
compactly supported test section in the dual vector bundle $V(i)^{\rm v}$ to
$V(i)$---to generate a *-algebra of observables, $\A(\M)$.
However, in most other algebraic approaches to quantum field theory,
$\A(\M)$ is assumed, a priori, to possess a particular topological
and/or other structure (e.g., C*-algebra structure) and the algebraic
relations within $\A(\M)$---together, perhaps, with specified actions
of symmetry groups on $\A(\M)$---are assumed to encode all of the
information about the quantum field theory under consideration. In
particular, since the state space, $\S(\M)$, is normally taken to
consist of all positive linear maps on $\A(\M)$, it is clear that
$\S(\M)$ cannot contain any information about the quantum field theory
that is not already contained in $\A(\M)$. We shall not proceed in
this manner because it is far from clear to us what topological and
other structure $\A(\M)$ should be assumed to possess a
priori in order to describe the quantum field theory. Instead, we shall
view the theory as being specified by
providing both an algebra of observables, $\A(\M)$, and a space of
allowed states, $\S(\M)$. Essentially the only information about the
theory contained in the algebra of obervables, $\A(\M)$, will be the
list of fields appearing in the theory and the relations that can be
written as polynomial expressions in the fields and their
derivatives. In our approach, the information normally encoded in the
topology of $\A(\M)$ will now be encoded in $\S(\M)$. Of course, the
semi-norms provided by $\S(\M)$ could be used, a postiori, to define a
topology on $\A(\M)$, but it is not clear that this topology would
encode all of the information in $\S(\M)$; in any case, we
find it simpler and more natural to consider the quantum field theory
to be defined by the pair $\{\A(\M), \S(\M)\}$.

\medskip

The key idea of this paper is that we will obtain the
pair $\{\A(\M), \S(\M)\}$
in a natural (i.e., functorial) manner from
the space of field labels $\I$ and
another datum, namely, the collection of ``operator product expansion (OPE)
coefficients''. The OPE coefficients are a family
\ben
\label{family}
\C(\M) \equiv \bigg\{ C^{(i_1) \cdots (i_n)}_{(j)}(x_1, \dots, x_n; y) \,\,\, :
\quad i_1, \dots, i_n, j \in \I, \,\, n \in {\mathbb N} \bigg\} \, ,
\een
where each $C^{(i_1) \cdots (i_n)}_{(j)}$ is a distribution on
$M^{n+1}$, valued in the vector bundle
\ben
E=V(i_1) \times \dots \times V(i_n) \times V(j)^{\rm v}
\stackrel{\pi}{\longrightarrow} M^{n+1}
\een
that is defined in some open
neighborhood of the diagonal in $M^{n+1}$. Thus, given $\C(\M)$,
we will construct both the algebra $\A(\M)$ and the state space $\S(\M)$
$$\begin{array}{rcc}
           \; & \;            & \S(M)      \\
           \; & \nearrow      & \mid    \\
           \C(\M)  & \;            & {\rm \tiny dual \, pairing}     \\
           \; & \searrow      & \mid     \\
           \; & \;            & \A(\M)
\end{array}$$
Thus, in our framework, a quantum field theory is
uniquely specified by providing a list of quantum fields
$\I$ and a corresponding list of OPE coefficients $\C(\M)$.
The OPE coefficients will be required to satisfy certain general
properties, which, in effect, become the ``axioms'' of quantum field
theory in curved spacetime. Most of the remainder of this
section will be devoted to formulating these axioms. However, before
providing the axioms for $\C(\M)$, we
briefly outline how the algebra $\A(\M)$
and the state space $\S(\M)$ are constructed from
$\C(\M)$ for any background structure
$\M = (M,g,T,e)$.

The {\em algebra} $\A(\M)$ is constructed by starting with the free
algebra ${\rm Free}(\M)$ generated by all expressions of the form
$\phi^{(i)}(f)$ with $i \in \I$, and $f$ a compactly supported test
section in the dual vector bundle to $V^{(i)}$.
We define an antilinear *-operation on ${\rm Free}(\M)$ by requiring that
its action on the generators be given by
\ben
[\phi^{(i)}(f)]^* = \phi^{(i^\star)}(\bar f) \, ,
\een
where $\bar f \in{\rm Sect}_0[\bar V(i)]$ is the conjugate test
section to $f\in {\rm Sect}_0(V{(i)})$. The *-algebra $\A(\M)$ is
taken to be the resulting free *-algebra factored by a 2-sided ideal
generated by a set of polynomial relations in the fields and their
derivatives. These relations consist of certain ``universal''
relations that do not depend on the particular theory under
consideration (such as linearity of $\phi^{(i)}(f)$ in $f$ and
(anti-)commutation relations) together with certain relations that
may arise from the OPE coefficients $\C(\M)$.  A precise enumeration of
the relations that define $\A(\M)$ will be given at the
beginning of section 3.

\medskip

The {\em state space} $\S(\M)$ is a subspace of the space of
all linear, functionals $\omega: \A(\M) \to \mc$
that are positive in the sense that
$\omega(A^* A) \equiv \langle A^* A \rangle_\omega \ge 0$ for
all $A \in \A(\M)$. This
subspace is specified as follows:
First, we require that for any state $\omega \in \S(\M)$, the
OPE coefficients in the collection $\C(\M)$ in eq.~\eqref{family}
appear in the expansion of the expectation value of the
product of fields $\langle \phi^{(i_1)}(x_1)
\cdots \phi^{(i_n)}(x_n)\rangle_\omega$ in terms of the
fields $\langle \phi^{(j)} (y)\rangle_\omega$
\ben
\label{OPEidea}
\left\langle\phi^{(i_1)}(x_1) \cdots \phi^{(i_n)}(x_n)\right\rangle_\omega \approx \sum_{j} C^{(i_1)
\dots (i_n)}_{(j)}(x_1, \dots, x_n; y) \, \left\langle \phi^{(j)}(y)
\right\rangle_\omega \, .
\een
Here ``$\approx$'' means that
this equation holds in a suitably strong sense
as an asymptotic relation in the limit that $x_1,
\cdots, x_n \rightarrow y$. A precise definition of what is meant by this
asymptotic relation
will be given in eq.~(\ref{OPEs}) below. Secondly, we require
that $\omega$ satisfy a microlocal spectrum condition that, in essence,
states that the singularities of $\langle \phi^{(i_1)}(x_1)
\dots \phi^{(i_n)}(x_n) \rangle_\omega$
are of ``positive frequency type'' in the
cotangent space $T^*_{(x_1, \dots, x_n)} M^n$. The precise
form of this condition will be formulated in terms of
the wave front set~\cite{hormander} (see eqs.~(\ref{gamtdef})
and~(\ref{msc}) below).

\medskip

We turn now to the formulation of the conditions that we shall impose
on the OPE coefficients $\C(\M)$. As indicated above, in our framework, these
conditions play the role
of axioms for quantum field theory.
Each operator product coefficient $C^{(i_1) \cdots (i_n)}_{(j)}$
in $\C(\M)$ is a distribution\footnote{
More precisely, each OPE coefficient is an equivalence class of distributions,
where two distributions are considered equivalent if their difference
satisfies eq.~(\ref{36}) below for all $\delta > 0$ and all $\T$.
Indeed, the OPE coefficients are more properly thought of as a sequence of
(equivalence classes of) distributions, such that the difference between the
$n$th and $m$th terms in the sequence satisfies eq.~(\ref{36})
for $\delta = {\rm min}(m,n)$. However, to avoid such an extremely cumbersome
formulation of our axioms and results, we will treat each OPE coefficient
as a distribution.}
on $M^{n+1}$,
valued in the vector bundle $V(i_1) \times \dots \times V(i_n) \times
V(j)^{\rm v}$ that is defined in some open neighborhood of the diagonal
in $M^{n+1}$. We will impose the following requirements
on these coefficients:

\begin{enumerate}
\item[C1)]  Locality and Covariance
\item[C2)]  Identity element
\item[C3)]  Compatibility with the $\star$-operation
\item[C4)]  Commutativity/Anti-Commutativity
\item[C5)]  Scaling Degree
\item[C6)]  Asymptotic positivity
\item[C7)]  Associativity
\item[C8)]  Spectrum condition
\item[C9)]  Analytic dependence upon the metric
\end{enumerate}

Before formulating these conditions in detail, for each $i \in \I$ we
define the {\it dimension}, $\d(i) \in \mr$, of the field $\phi^{(i)}$
by\footnote{Note that, when $V(i)$ is not equal to $M\times \mc$, i.e., when
$\phi^{(i)}$ is not a scalar field, then quantities like $C^{(i)(i^\star)}_{(\myid)}$
are, by definition, distributions taking values in a vector bundle. What we mean by
the scaling degree here and in similar equations in the following such as eq.~\eqref{sdcijk}
is the maximum of the scaling degrees of all "components"
of such a bundle-valued distribution.}
\ben\label{dimdef}
\d(i) := \frac{1}{2} \sup_{{\rm backgrounds} \,\, \M} \sd \bigg\{ C^{(i)(i^\star)}_{(\myid)} \bigg\} \, ,
\een
where "sd" denotes the scaling degree of a distribution (see appendix A)
and it is understood that the scaling degree is taken about a
point on the diagonal.
In other words, $\d(i)$ measures the rate at which the
coefficient of the identity $\myid$ in the operator product expansion of
$\phi^{(i)}(x_1) \phi^{(i^\star)}(x_2)$ blows up as $x_1 \rightarrow x_2$.
It will follow immediately from
condition (C3) below that $\d(i^\star) = \d(i)$.
Note also that $\d(\myid) = 0$.

For distributions $u_1$ and $u_2$ on $M^{n+1}$, we introduce the equivalence
relation
\ben\label{simdef}
u_1 \sim u_2
\een
to mean that the scaling degree of the
distribution $u = u_1 - u_2$ about any
point on the total diagonal is $-\infty$. However, it should
be noted that in the formulation of
condition (C8) below and in the precise definition of the operator
product expansion, eq.~(\ref{OPEs}), we will need to consider limits
where different
points approach the total diagonal at different rates. We will then
introduce a stronger notion of equivalence that, in effect, requires
a scaling degree of $-\infty$ under
these possibly different rates of approach. It is this stronger
notion that was meant in eq.~(\ref{OPEidea}) above.

\medskip
\noindent
\paragraph{\bf (C1) Covariance:}
Let $\rho: \M \to \M'$ be a causality preserving isometric embedding preserving
the orientations, spin-structures, and all other background structure, i.e.
$\rho$ is a morphism in the category of background structures.
We postulate that
for each member of the above collection~\eqref{family}, we have
\ben
(\rho^* \times \cdots \rho^* \times \rho^{-1}_*) \,
C^{(i_1) \dots (i_n)}_{(j)}[\M'] \sim C^{(i_1) \dots (i_n)}_{(j)}[\M] \, .
\een

\medskip
\noindent
\paragraph{\bf (C2) Identity element:}
We require that
\ben
C^{(i_1) \dots (\myid) \dots (i_{n})}_{(j)}(x_1, \dots, x_{n}; y) =
C^{(i_1) \dots (i_{k})(i_{k+1})
  \dots (i_{n})}_{(j)}(x_1, \dots x_{k-1}, x_{k+1}, \dots x_n; y) \, ,
\een
where the identity $\myid$ is in the $k$-th place. We also impose
the following additional conditions on
$C^{(i)}_{(j)}(x; y)$, since these coefficients
should merely implement a Taylor expansion
of the operator with label $i$ localized at $x$ in terms of
operators with labels $j$
localized at point $y$. As in a Taylor series,
we demand that these
coefficients depend only polynomially on the Riemannian normal
coordinates of $x$ relative to $y$
(and are thus in particular smooth), and that
\ben
C^{(i)}_{(j)}(x;x)
= \delta^{(i)}_{(j)} \, {\rm id}_{(i)}^{}(x) \, ,
\een
where ${\rm id}_{(i)}(x)$ is the identity map in the fiber over $x$
of the vector bundle $V(i)$.
Since a Taylor expansion of an operator at $x$ around another point $y$ involves
the derivatives of the operators considered at $y$, and because derivatives
tend to increase the dimension, we further demand that $C^{(i)}_{(j)} = 0$ if
${\rm dim}(j) < {\rm dim}(i)$.
Finally, if we Taylor
expand a quantity at $x_1$ successively around a second point $x_2$,
and then a third point $x_3$, this should be equivalent
to expanding it in one stroke around the third point. Thus, we require that we have
\ben
C^{(i)}_{(j)}(x_1; x_3) = \sum_{(k)}  C^{(i)}_{(k)}(x_1; x_2) C^{(k)}_{(j)}(x_2; x_3) \, ,
\een
where we note that the sum is only over the (finitely many) field labels $k$ such
that ${\rm dim}(k) \le {\rm dim}(j)$.

\medskip
\noindent
\paragraph{\bf (C3) Compatibility with $\star$:}
This relation encodes the fact that the underlying theory will have
an operation analogous to the hermitian adjoint of a linear operator.
The requirement is
\ben
\overline{C^{(i_1) \dots (i_n)}_{(j)}(x_1, \dots, x_n, y)} \sim
C^{(i_n^\star) \dots (i_1^\star)}_{(j^\star)} \circ \pi(x_1, \dots,
x_n, y) \, ,
\een
where $\pi$ is the permutation
\ben\label{permutationpi}
\pi =
\left(
\begin{matrix}
1 & 2 & \dots & n & n+1\\
n & n-1 & \dots & 1 & n+1
\end{matrix}
\right) \, .
\een

\medskip
\noindent
\paragraph{\bf (C4) Commutativity/Anti-Commutativity:}
Let $\sigma$ be the permutation
\ben\label{sigmadef}
\sigma =
\left(
\begin{matrix}
1 & \dots & k & k+1 & \dots & n+1\\
1 & \dots & k+1 & k & \dots & n+1
\end{matrix}
\right) \, ,
\een
and let $F(i)$ be the Bose/Fermi parity of $\phi^{(i)}$.
Then we have
\ben\label{causal}
C^{(\sigma i_1) \dots (\sigma i_n)}_{(j)} \circ \sigma (x_1, \dots,
x_n, y)
= -(-1)^{F(i_k)F(i_{k+1})} C^{(i_1) \dots (i_n)}_{(j)} (x_1, \dots, x_n, y) \,
\een
whenever $x_k$ and $x_{k+1}$ are spacelike separated (and in the
neighborhood of the diagonal in $M^{n+1}$
where the OPE coefficients are actually defined).

\medskip
\noindent
\paragraph{\bf (C5) Scaling Degree:}

We require that
\ben\label{sdcijk}
\sd \bigg\{ C^{(i)(j)}_{(k)} \bigg\} \le \d(i)+\d(j)-\d(k) \, .
\een

\medskip
\noindent
\paragraph{\bf (C6) Asymptotic Positivity:}
Let $i \in \I$ be any given index. Then, for $D \ge 3$ we postulate
that $\d(i)  \ge 0$, and that $\d(i) = 0$ if and only if $i=\myid$.
Note that, because we are
taking the supremum over all spacetimes in eq.~\eqref{dimdef}, our requirement that
$\d(i) > 0$ for $i \neq \myid$ does not imply
that the scaling degree of $C^{(i)(i^\star)}_{(\myid)}$ for $i \neq \myid$
is positive for {\em all} spacetimes, since the coefficient may e.g.
``accidentally'' happen to have a lower scaling degree for certain spacetimes of
high symmetry, as happens for certain supersymmetric theories on Minkowski spacetime.

On a spacetime $M$ where the scaling degree of $C^{(i)(i^\star)}_{(\myid)}$
is $\d(i)$, we know that if we scale the arguments of this distribution
together by a factor of $\lambda$, and multiply by a power of $\lambda$ less than $2\d(i)$,
then the resulting family of distributions cannot be
bounded as $\lambda \to 0$. For our applications below, it is convenient
to have a slightly stronger property, which we now explain.
Let $X^a$ be a vector field on $M$ locally defined near
$y$ such that $\nabla_a X^b  = -\delta_a{}^b$ at $y$.
Let $\Phi_t$ be the flow of this field, which
scales points by a factor of $e^{-t}$ relative to
$y$ along the flow lines of $X^a$.
If $f$ is a compactly supported test section in $V(i)$,
we set $f_\lambda = \lambda^{-D} \, \Phi^*_{\log \lambda} f$.
This family of test sections becomes more and more sharply peaked at
$y$ as $\lambda \to 0$. We postulate that, for any $\delta>0$ and
any $X^a$ as above, there exists an $f$ such that
\ben\label{apc}
\lim_{\lambda \to 0}  \bigg| \lambda^{2\d(i)-\delta}
\int_{M \times M} C^{(i)(i^\star)}_{(\myid)}(x_1, x_2,y)
f_\lambda(x_1)
\bar f_\lambda(x_2) \,
d\mu_1 d\mu_2 \bigg| =  \infty \, ,
\een
uniformly in $y$ in some neighborhood. This statement is slightly stronger than the
statement that the scaling degree of our distribution is $\d(i)$ on $M$,
since the latter would only imply that the rescaled distributions under the
limit sign in eq.~\eqref{apc} contain a subsequence that is unbounded in $\lambda$
for some test section $F(x_1, x_2)$ in $V(i) \times V(i^\star)$, not necessarily
of the form $f(x_1) \bar f(x_2)$.

The reason for the terminology
``asymptotic positivity axiom'' arises from
lemma~\ref{scalelemma} below. An alternative essentially equivalent
formulation of this condition, which is related to ``quantum inequalities'', is
given in Appendix~B.

In $D=2$ spacetime dimensions,
the above form of the asymptotic positivity condition is in general too
restrictive. The reason is that in $D=2$, there are usually many
fields $\phi^{(i)}$ of dimension $\d(i) = 0$ different from the identity operator.
For example, for a free Klein-Gordon field, the basic field $\varphi$ and
all its Wick-powers have vanishing dimension---in fact,
their OPE-coefficients have a logarithmic scaling behavior. A possible way to
deal with this example would be to consider only composite fields containing
derivatives, as this subspace of fields is closed under the OPE. For
this subspace of fields, the asymptotic positivity condition would then
hold as stated. Another possibility is to introduce a suitably refined measure
of the degree of divergence of the OPE coefficients also taking into account
logarithms. Such a concept would clearly be sensible for
free or conformal field theories in $D=2$, and it would also be
adequate in perturbation theory (to arbitrary but finite orders).
A suitable refinement of the above asymptotic positivity condition
could then be defined, and all proofs given in the remainder
of this paper would presumably still hold true, with minor modifications.
For simplicity, however, we will not discuss this issue further in this paper,
and we will stick with the asymptotic positivity condition in the above form.

\medskip
\noindent
\paragraph{\bf (C7) Spectrum condition:}

The spectrum condition roughly says that the singularities of
a field product ought to be of ``positive frequency type,''
and is completely analogous the condition imposed on
states that we will impose below: We demand that, near the diagonal,
the wave front set (see Appendix~\ref{appwfs}) of the OPE coefficient satisfies
\ben
\WF(C^{(i_1) \dots (i_n)}_{(j)}) \subset \Gamma_n(\M) \times Z^*M \, ,
\een
where the last factor
$Z^*M$ is the zero section of $T^*M$ and corresponds to the reference
point $y$ in the OPE, and where
the set $\Gamma_n(\M) \subset T^*M^n \setminus \{0\}$ is defined
as follows. Consider embedded graphs $G(\x, \vec y, \vec p) \in {\mathcal
  G}_{m,n}$ in the spacetime manifold $M$ which have the following
properties. Each graph $G$ has $n$ so-called ``external vertices'',
$x_1, \dots, x_n \in M$, and $m$ so-called ``internal'' or
``interaction vertices'' $y_1, \dots, y_m \in M$. These vertices are
of arbitrary valence, and are joined by edges, $e$, which are
null-geodesic curves $\gamma_e: (0,1) \to M$. It is assumed that
an ordering of the vertices is defined, and that the ordering
among the external vertices is $x_1 < \dots < x_n$, while the
ordering of the remaining interaction vertices is unconstrained.
If $e$ is an edge joining two vertices, then $s(e)$ (the source) and
$t(e)$ (the target) are the two vertices $\gamma_e(0)$ and
$\gamma_e(1)$, where the curve
is oriented in such a way that it starts at the smaller vertex
relative to the fixed vertex ordering. Each edge carries a future
directed, tangent parallel covector field, $p_e$,
meaning that $\nabla_{\dot \gamma_e} p_e = 0$, and $p_e \in \partial
V^+$. With this notation set up, we define
\begin{eqnarray}
\label{gamtdef}
\Gamma_{n,m}(M, g) &=&
\Bigg\{(x_1, k_1; \dots; x_n, k_n) \in T^*M^n \setminus \{0\} \mid
\exists \,\, \text{decorated graph $G(\x, \vec y, \vec p) \in
  {\mathcal G}_{m,n}$} \nonumber\\
\vspace{0.3cm}
&& \text{such that $y_i \in J^+(\{x_1, \dots, x_n\}) \cap J^-(\{x_1,
  \dots, x_n\})$ for all $1 \le i \le m$,} \nonumber\\
\vspace{0.3cm}
&& \text{such that
$k_i = \sum_{e: s(e) = x_i} p_e - \sum_{e: t(e) = x_i} p_e$ for all
$x_i$ and}\nonumber\\
&& \text{such that
$0 = \sum_{e: s(e) = y_i} p_e - \sum_{e: t(e) = y_i} p_e$ for all
$y_i$}\Bigg\} \, ,
\end{eqnarray}
where $J^\pm(U)$ is the causal future resp. past of a set $U \subset M$, defined
as the set of points that can be reached from $U$ via a future
resp. past directed causal curve. We set
\ben
\label{msc}
\Gamma_n = \overline{\bigcup_{m \ge 0} \Gamma_{m,n}} \, .
\een
Note in particular that the microlocal spectrum condition implies that the
dependence of our OPE coefficients~\eqref{family} on the reference point $y$ is
smooth. When the spacetime is real analytic, we require a similar
condition for the "analytic" wave front set~\cite{hormander} $\WF_A$ of
the OPE coefficient.

Our formulation of the microlocal
spectrum condition is a weaker condition than
that previously proposed in~\cite{bfk},
based on earlier work of~\cite{radzikowski}. The microlocal
spectrum condition of~\cite{bfk} is satisfied by the
correlation functions of suitable Hadamard states in linear field theory,
but need not hold even perturbatively for interacting fields.
In essence, our formulation allows for the presence of interaction vertices,
thus weakening the condition relative to the free field case.
Our condition can be shown to hold for perturbative interacting
fields~\cite{hollands2006}.

\medskip
\noindent
\paragraph{\bf (C8) Associativity:}
Following~\cite{hollands2006},
a notion of associativity is formulated by considering
configurations $(x_1, \dots, x_n)$ of points in $M^n$ (where $n>2$)
approaching a point $y \in M$ at different rates.
For example, if we have 3 points $(x_1, x_2, x_3)$, we may
consider all points coming close to each other at the same rate, i.e., assume
that their mutual distances are of order $\varepsilon$, where $\varepsilon
\to 0$. Alternatively, we may consider a situation in which,
say, $x_1, x_2$ approach each other very fast,
say, at rate $\epsilon^2$, while $x_3$ approaches
$x_1, x_2$ at a slower rate, say at rate $\epsilon$.
The first situation corresponds,
intuitively, to the process of performing the OPE of
a triple product of operators ''at once'', while the second
situation corresponds to
first performing an OPE in the factors corresponding to $x_1, x_2$,
and then successively
performing a second OPE between the resulting fields and the
third field situated
at $x_3$. Obviously, for an arbitrary number $n$ of points,
there are many different
possibilities in which configurations can come close.
We classify the different possibilities in terms of ``merger trees,'' $\T$.
Each merger tree will give rise to a separate associativity condition.

For this, one constructs curves
in $M^n$ parametrized by $\epsilon$,
which are in $M^n_0$ (the space $M^n$ minus all its
diagonals) for $\epsilon > 0$, and which tend to a point
on the diagonal as $\epsilon \to 0$.
These curves are labeled by trees $\T$ that characterize
the subsequent mergers of the points in
the configuration as $\epsilon \to 0$. A convenient way to formally
describe a tree $\T$ (or more generally, the disjoint union of trees,
a ``forest'') is by a nested set $\T = \{S_1, \dots, S_k\}$ of subsets
$S_i \subset \{1, \dots, n\}$. ``Nested'' means that two sets are
either disjoint, or one is a proper subset of the other.
We agree that the sets $\{1\}, \dots, \{n\}$ are always contained in
the tree (or forest). Each set $S_i$ in $\T$ represents a node of a
tree, i.e., the set of vertices ${\rm Vert}(\T)$ is given by the
sets in $\T$, and $S_i \subset S_j$ means
that the node corresponding to $S_i$ can be reached by moving downward from the
node represented by $S_j$. The root(s) of the tree(s) correspond to the
maximal elements, i.e., the sets that are not subsets of any other
set. If the set $\{1, \dots, n\} \in \T$, then there is in fact only
one tree, while if there are several maximal elements, then there are
several trees in the forest, each maximal element corresponding to
the root of the respective tree. The leaves
correspond to the sets $\{1\}, \dots, \{n\}$, i.e., the minimal elements.
For example for a configuration of $n=4$ points, a tree might look like in
the following figure, and the corresponding nested set of subsets is also given.
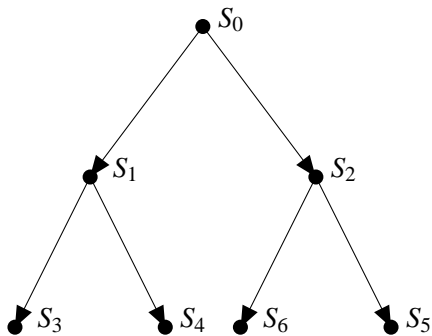
\setlength{\unitlength}{1cm}
\begin{center}
\begin{picture}(6,6.5)(0.0,0.0)
\put(3,6){\circle*{0.2}}
\put(3.2,6){$S_0$}
\put(1.5,4){\circle*{0.2}}
\put(1.8,4){$S_1$}
\put(4.5,4){\circle*{0.2}}
\put(4.7,4){$S_2$}
\put(0.5,2){\circle*{0.2}}
\put(0.8,2){$S_3$}
\put(2.5,2){\circle*{0.2}}
\put(2.7,2){$S_4$}
\put(5.5,2){\circle*{0.2}}
\put(5.7,2){$S_5$}
\put(3.5,2){\circle*{0.2}}
\put(3.8,2){$S_6$}
\put(3,6){\vector(-1.5,-2){1.5}}
\put(3,6){\vector(1.5,-2){1.5}}
\put(1.5,4){\vector(-1,-2){1}}
\put(1.5,4){\vector(1,-2){1}}
\put(4.5,4){\vector(1,-2){1}}
\put(4.5,4){\vector(-1,-2){1}}
\put(1.2,1){${\bf T} = \{S_0, S_1, \dots, S_6\}$}
\put(-4,0.5){$S_0 = \{1, 2, 3, 4\}, \,\, S_1 = \{1, 2\}, \,\, S_2 =
  \{3, 4\}, \,\, S_3=\{1\}, \,\, S_4=\{2\}, \,\, S_5=\{3\}, \,\, S_6=\{4\}$}
\end{picture}
Figure~1.
\end{center}

In the following, we will consider only $\T$ with a single
root. The desired curves $\x(\epsilon)$ tending to the diagonal
are associated with $\T$ and are constructed as
follows. First, we construct Riemannian normal coordinates
around the reference point $y$, so that each point in a
convex normal neighborhood of $y$ may be identified with a
tangent vector $v \in T_{y} M$. We then choose a tetrad and further
identify $T_{y} M \cong \mr^D$, so that $v$ is in fact viewed as
an element in $\mr^D$. With each set $S \in \T$, we now associate
a vector $v_S \in \mr^D$, which we collect in a tuple
\ben
\vec v = (v_{S_1}, \dots, v_{S_r}) \in (\mr^D)^{|\T|}
\,, \quad \T = \{S_1, \dots, S_r\}
\, ,
\een
and we agree that $v_{\{1, \dots, n\}}=0$, and 
where $|\T|$ is the number of nodes of the tree, i.e. 
the number of elements of the set $\T$. For $\varepsilon > 0$, we
define a mapping
\ben
\psi_\T(\varepsilon): (\mr^D)^{|\T|} \mapsto (\mr^D)^{|\T|},
(v_{S_1}, \dots, v_{S_r}) \mapsto (x_{S_1}(\varepsilon), \dots,
x_{S_r}(\varepsilon))
\een
by the formula
\ben
x_S(\varepsilon) = \sum_{S' \subset S} \epsilon^{depth(S')} v_{S'}
\een
where $depth(S')$ is defined as the number of nodes that connect $S'$
with the root of the tree $\T$. For $\epsilon$ sufficiently small, and
$\vec v$ in a ball $B_1(0)^{| \T |}$, the vectors $x_S(\varepsilon) \in
\mr^D$ may be identified with points in $M$ via the exponential map.
If the vectors $v_S$ satisfy the condition that, $v_{S'} \neq v_{S''}$
for any $S', S''$ that are connected to a common $S$ by an edge,
then the vector $(x_{\{1\}}(\varepsilon), \dots, x_{\{n\}}(\varepsilon)) \in M^n$ does
not lie on any of the diagonals, i.e., any pair of entries are
distinct from each other. Its value as $\varepsilon \to 0$ approaches
the diagonal of $M^n$. The $i$-th point in the configuration
$x_{\{i\}}(\epsilon)$ is obtained starting from $y$
by following the branches of the tree towards the $i$-th leaf,
moving along the first edge by an amount $\epsilon$
in the direction of the corresponding $v_S$
then the second by an amount $\epsilon^2$ in the direction of
the corresponding $v_S$,
and so fourth, until the $i$-th leaf is reached.
The curve
$(x_{\{1\}}(\varepsilon), \dots, x_{\{n\}}(\varepsilon)) \in M^n$
thus
represents a configuration of points which merge hierarchically according to
the structure of the tree $\T$, as $\epsilon \to 0$. That is,
the outermost branches of the tree merge at the highest
order in $\epsilon$, i.e., at rate $\epsilon^{\rm depth \, of \, branch}$,
then the next level at a lower order, and so fourth, while
the branches closest to the root merge at the slowest rate, $\epsilon$.
The following figure illustrates our definition.
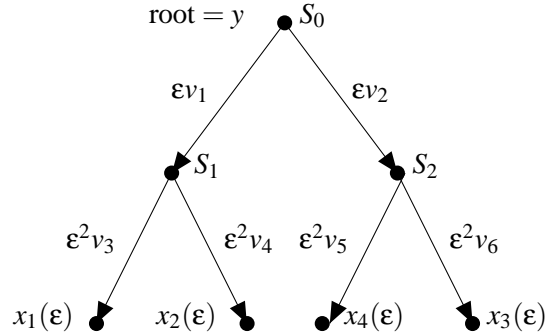
\setlength{\unitlength}{1cm}
\begin{center}
\begin{picture}(6,6.5)(0.0,0.0)
\put(3,6){\circle*{0.2}}
\put(3.2,6){$S_0$}
\put(1.2,6){${\rm root}=y$}
\put(1.5,4){\circle*{0.2}}
\put(1.8,4){$S_1$}
\put(4.5,4){\circle*{0.2}}
\put(4.7,4){$S_2$}
\put(0.5,2){\circle*{0.2}}
\put(-0.6,2){$x_1(\epsilon)$}
\put(2.5,2){\circle*{0.2}}
\put(1.3,2){$x_2(\epsilon)$}
\put(5.5,2){\circle*{0.2}}
\put(5.7,2){$x_3(\epsilon)$}
\put(3.5,2){\circle*{0.2}}
\put(3.8,2){$x_4(\epsilon)$}
\put(3,6){\vector(-1.5,-2){1.5}}
\put(1.5 ,5){$\epsilon v_1$}
\put(3,6){\vector(1.5,-2){1.5}}
\put(3.9 ,5){$\epsilon v_2$}
\put(1.5,4){\vector(-1,-2){1}}
\put(0.1 ,3){$\epsilon^2 v_3$}
\put(1.5,4){\vector(1,-2){1}}
\put(2.2,3){$\epsilon^2 v_4$}
\put(4.5,4){\vector(1,-2){1}}
\put(3.2 ,3){$\epsilon^2 v_5$}
\put(4.6,4){\vector(-1,-2){1}}
\put(5.2 ,3){$\epsilon^2 v_6$}
\put(1.2,1){${\bf T} = \{S_0, S_1, \dots, S_6\}$}
\put(-4,0.5){$x_1(\epsilon)=\epsilon v_1+\epsilon^2 v_3, \,\,
x_2(\epsilon)=\epsilon v_1 + \epsilon^2 v_4, \,\, x_3(\epsilon)= \epsilon
v_2 + \epsilon^2 v_5, \,\, x_4(\epsilon)= \epsilon v_2 + \epsilon^2 v_6$}
\end{picture}
Figure~2
\end{center}
Thus, the points are scaled towards the diagonal of $M^n$, even though
possibly at different speeds, and the limiting element as
$\varepsilon \to 0$ is the element $(y, \dots, y)$ on the diagonal\footnote{
However, if the vector
$(x_{\{1\}}(\varepsilon), \dots, x_{\{n\}}(\varepsilon)) \in M^n$
is alternatively viewed
as an element of the ``Fulton-MacPherson compactification'' $M^n_c$ of the
configuration space $M^n$, then its limiting value may be viewed
alternatively as lying in the boundary $\partial M^n_c$
of the compactification, and
the vectors $\vec v$ may be viewed as defining a coordinate system
of that boundary, which thereby has the structure of a stratifold
\ben
\partial M^n_c \cong \bigcup_{\T} M[\T] \, ,
\een
with each face $\T$ corresponding to a lower dimensional subspace
associated with a given merger tree~\cite{Singer1990}.}.

\medskip
Using the maps $\psi_\T(\epsilon)$ we can define an
asymptotic equivalence relation $\sim_{\T, \delta}$ for
distributions $u$ defined on $M^{|\T|}$. For
points within a convex normal neighborhood, and sufficiently small
$\varepsilon > 0$, we can define the pull-back
$u \circ \psi_\T(\varepsilon)$. This may be viewed a distribution in
the variables $v_S \in \mr^D, S \in \T$.
We now define
\ben\label{36}
u \sim_{\T, \delta} 0 \quad :\Longleftrightarrow \quad
\lim_{\epsilon \to 0+} \epsilon^{-\delta} \,
u\circ \psi_\T(\epsilon)
= 0 \, \quad {\rm for \,\,\,}
\delta>0 \, ,
\een
in the sense of distributions defined on a neighborhood of the origin
in $(\mr^D)^{|\T|}$. We write $u \approx 0$ if $u \sim_{\T, \delta} 0$
for all $\T$ and all $\delta$. The condition that $u \approx 0$ is
stronger than  the
previously defined condition $u \sim 0$ [see eq.~\eqref{simdef}], which corresponds to the
requirement that $u \sim_{\T, \delta} 0$ for all $\delta$
only for the trivial tree $\T = \{\{1\},\dots \{n\},\{1, \dots, n\}\}$.

\medskip

We can now state the requirement of associativity.
Recall that if $\T$ is a tree with $n$
leaves, then $\psi_\T(\epsilon)$ gives a
curve in the configuration space of
$n$ points in $M$ which representing the process of a
subsequent hierarchical merger of the
points according to the structure of the tree. If a subset
of points in $(x_{\{1\}}(\epsilon), \dots, x_{\{n\}}(\epsilon))$
merges first, then one intuitively expects that one should be able
to perform the OPE in those points first, and then
subsequently perform OPE's of the other points
in the hierarchical order represented by the tree. We will impose this
as the associativity requirement. For example, if
we have 4 points, and the tree corresponds to the nested set of subsets
$\T=\{\{1,2\}, \{3,4\}, \{1,2,3,4\}\}$ as in the above figure,
the pairs of points $x_{\{1\}}(\epsilon), x_{\{2\}}(\epsilon)$ respectively
$x_{\{3\}}(\epsilon), x_{\{4\}}(\epsilon)$
approach each other at order $\epsilon^2$, while the two groups
then approach each other at a slower rate $\epsilon$. We postulate that\footnote{
Here, the distribution on the left side is viewed as a distribution in
$x_1, \dots, x_7, y$ with a trivial dependence on $x_5, x_6, x_7$.}
\begin{multline}\label{38}
C^{(i_1) (i_2) (i_3) (i_4)}_{(i_5)}(x_1, x_2, x_3, x_4; y) \sim_\T
\\ \sum_{i_6, i_7}
C^{(i_1) (i_2)}_{(i_6)}(x_1, x_2; x_6) \, C^{(i_3) (i_4)}_{(i_7)}(x_3,
x_4; x_7) \,
C^{(i_6) (i_7)}_{(i_5)}(x_6, x_7; y) \, .
\end{multline}
For the same product of operators, consider alternatively the tree
$\T' = \{\{1,2,3\}, \{1,2,3,4\}\}$. The corresponding
associativity relation for the OPE coefficient is now
\ben\label{39}
C^{(i_1) (i_2) (i_3) (i_4)}_{(i_5)}(x_1, x_2, x_3, x_4; y) \sim_{\T'}
\sum_{i_6} C^{(i_1) (i_2) (i_3)}_{(i_6)}(x_1, x_2, x_3; x_6) \,
C^{(i_6) (i_4)}_{(i_5)} (x_6, x_4; y)
\een
It is important to note, however, that there is in general no simple
relation between the right hand sides of eqs.~\eqref{38},\eqref{39}
for different trees $\T$ and $\T'$.

The corresponding relation for arbitrary numbers of points,
and arbitrary types of trees is a straightforward generalization of this case,
the only challenge being to introduce an appropriate notation to express the
subsequent OPE's. For this, we consider maps
$\vec i: \T \to \I$ which associate with
every node $S \in \T$ of the tree an element $i_S \in \I$, the index
set labelling the fields. If $S \in \T$, we let
$S(1), S(2), \dots S(r)$ be the branches of this tree, i.e. the nodes
connected to $S$ by a single upward edge.
With these notations in place, the generalization of eqs.~\eqref{38}
and~\eqref{39} for an
arbitrary number $n$ of points, and an arbitrary tree $\T$
is as follows. Let $\T$ be an arbitrary tree on $n$ elements,
and let $\delta > 0$ be an arbitrary real number. Then we have\footnote{Here, the distribution on the left
is regarded as a distribution in $x_S; S \in \T$, with a
trivial dependence on the $x_S$ with $S$ not equal
to $\{1\}, \dots, \{n\}, \{1, \dots, n\}$.}
\begin{equation}
\label{41}
C^{(i_1) \dots (i_n)}_{(j)}(x_1, \dots, x_n; y) \,\,
\sim_{\T,\delta} \,\,
\sum_{\vec i \in {\rm Map}(\T, I)} \left( \prod_{S \in \T}
C^{ (i_{S(1)}) \dots (i_{S(r)})}_{(i_S)}
\left( x_{S(1)}, \dots, x_{S(r)}; x_{S}\right) \right) \, ,
\end{equation}
where the sums are over $\vec i$ with the properties that
\ben
i_{\{1\}} = i_1, \dots, i_{\{n\}} = i_n \,, \quad i_{\{1, \dots, n\}} = j \, .
\een
The sum over $\vec i$ is finite, with $\d(i_S) \le \Delta$, where $\Delta$
is a number depending on the tree $\T$ and the real number $\delta$. Furthermore,
it is understood that $x_{\{1\}} = x_1, \dots, x_{\{n\}} = x_n$, and that $x_{\{1, \dots, n\}} = y$.

\medskip
\noindent
\paragraph{\bf (C9) Analytic and smooth dependence:} Due to requirement (C1),
the OPE coefficients may be regarded as functionals of the
spacetime metric. We require that the
distributions $C^{(i_1) \dots (i_n)}_{(j)}$ have an analytic dependence
upon the spacetime metric. For this, let $g^{(s)}$ be a
1-parameter family of analytic metrics, depending analytically on $s \in \mr$.
Then the corresponding OPE-coefficients $C^{(i_1) \dots (i_n)}_{(j)}$
are distributions in $x_1, \dots, x_n, y$ that also depend
on the parameter $s$.
We demand that the dependence on $s$ is "analytic". It is technically
somewhat involved to define what one precisely means by this, because
$C^{(i_1) \dots (i_n)}_{(j)}$ itself is not analytic, but
instead a distribution in the spacetime points. The appropriate way
to make this definition rigorous was provided in~\cite{hw2,hpct}. Similarly,
if the spacetime is only smooth, we require a corresponding smooth
variation of the OPE coefficients under smooth variations of the metric.


\section{Construction of the QFT from the OPE coefficients}\label{section3}

Now that we have stated in detail all the desired properties of the
OPE coefficients, we are ready to give the precise definition
of quantum field theory.
A quantum field theory in curved spacetime associated with a collection of OPE coefficients
satisfying the above properties is the pair $\{\A(\M), \S(\M)\}$
consisting of a *-algebra $\A(\M)$ and a space of states
$\S(\M)$ on $\A(\M)$ that is canonically
defined by the operator product coefficients, $\C(\M)$,
for any choice of the background spacetime structure $\M$.
The algebra $\A(\M)$ is defined as follows. To begin,
let ${\rm Free}(\M)$ be the free
*-algebra generated by all expressions of the form $\phi^{(i)}(f)$ with $i \in \I$, and
$f$ a compactly supported test section in the vector bundle $V{(i)}$ associated with the
tensor or spinor character of $\phi^{(i)}$. The algebra $\A(\M)$ is obtained by factoring
${\rm Free}(\M)$ by a set of relations, which are as follows.

\medskip
\noindent
\paragraph{A1) \bf Identity:} We have $\phi^{(\myid)}(f) = \int f \,
d\mu \cdot \myid$.

\medskip
\noindent
\paragraph{A2) \bf Linearity:} For any complex numbers $a_1, a_2$, any test sections
$f_1, f_2$, and any field $\phi^{(i)}$, we have $\phi^{(i)}(a_1 f_1 + a_2 f_2) = a_1 \phi^{(i)}(f_1) + a_2 \phi^{(i)}(f_2)$.
The linearity condition might be viewed as saying that, informally,
\ben
\phi^{(i)}(f) = \int_M \phi^{(i)}(x) f(x) d\mu
\een
is a pointlike field that averaged against a smooth weighting function. We shall often use the informal pointlike fields as a notational
device, with the understanding that all identities are
supposed to be valid after formally smearing with a test function.

\medskip
\noindent
\paragraph{A3) \bf Star operation:} For any field $\phi^{(i)}$, and any test section $f\in {\rm Sect}_0(V{(i)})$,
let $\bar f \in{\rm Sect}_0(\bar V{(i)})$, be the conjugate test section. Then we require that
\ben\label{starop}
[\phi^{(i)}(f)]^* = \phi^{(i^\star)}(\bar f) \, .
\een

\medskip
\noindent
\paragraph{A4) \bf Relations arising from the OPE:} Let ${K} \subset \I$ be a subset
of the index set, and let $c_{(i)}, i \in {K}$ be scalar valued differential operators
[i.e., differential operators taking a section $V(i)$ to a scalar function on $M$],  such that
\ben\label{sdreq}
\sd \bigg\{ \sum_{i,j \in {K} } (\bar c_{(i^\star)} \otimes c_{(j)}) C^{(i^\star)(j)}_{(k)} v^{(k)} \bigg\}
< 0 \, ,
\een
for all $v^{(k)} \in V(k)$ and all $k \in \I$.
Then we impose the relation
\ben\label{Odef}
0 = \sum_{i \in {K}} c_{(i)} \phi^{(i)}(f)
\een
for all $f \in C_0^\infty(M)$, where the differential operators
act in the sense of distributions.
This relation can be intuitively understood as follows. Let $\O(f)$
be the smeared quantum field defined by
the right side of eq.~\eqref{Odef}. If we consider the OPE of the quantity
$\langle \O(x_1) \O(x_2)^* \rangle_\omega$ in some state,
then the scaling degree requirement~\eqref{sdreq}
implies that the limit of this quantity as $x_1, x_2 \to y$
is equal to 0. Heuristically, this implies that
$\O(y)$ is a well-defined element at a {\em sharp} point $y$,
not just after smearing with a test function
$f$ as in eq.~\eqref{Odef}. Since
$\langle \O(y) \O(y)^* \rangle_\omega = 0$ in all states, one would heuristically
conclude that also $\O(y)$ should be 0 as an
algebra element. This is what our requirement states.

The above requirement
serves to eliminate any redundancies in the field content arising e.g.
from initially viewing, say, a field $\varphi$ and
$\square \varphi$ as independent fields, or from initially
specifying a set of linearly dependent fields.
More nontrivially, this requirement
should also serve to impose field equations in $\A(\M)$.
For example, in $\lambda \varphi^4$-theory, we expect that
a field equation of the form
$\square \varphi - m^2 \varphi - \lambda \varphi^3 = 0$ should hold,
where $\varphi^3$ is a composite field in the theory
that should appear in the operator product expansion
of three factors of $\varphi$. If such a field equation holds, then
clearly $\square \varphi - m^2 \varphi - \lambda \varphi^3$
should have a trivial OPE with itself, i.e., all
OPE coefficients of the product of
this operator with itself should have arbitrary low scaling degree.
Thus, in this example, if we take
$c_{(1)} = \square - m^2, c_{(2)} = -\lambda$, and
$\phi^{(1)} = \varphi, \phi^{(2)} = \varphi^3$, then eq.~(\ref{sdreq})
should hold.
Our requirement effectively demands that field equations hold
if and only if they are implied by the OPE condition eq.~(\ref{sdreq}).

\medskip
\noindent
\paragraph{A5) \bf (Anti-)commutation relations:} Let $\phi^{(i_1)}$ and $\phi^{(i_2)}$ be fields, and let
$f_1$ and $f_2$ be test sections corresponding to their respective spinor or tensor character, whose supports are assumed
to be spacelike separated. Then the relation
\ben\label{comm}
\phi^{(i_1)}(f_1) \phi^{(i_2)} (f_2) +
(-1)^{F{(i_1)} F{(i_2)}} \phi^{(i_2)} (f_2) \phi^{(i_1)} (f_1) = 0 \, .
\een
holds in $\A(\M)$.

\vspace{1cm}

Having defined the algebra $\A(\M)$, we next define the
state space $\S(\M)$ to consist of all those linear functionals
$\langle \, . \, \rangle_\omega:
\A(\M) \to \mc$ with the following properties:

\medskip
\noindent
\paragraph{S1) \bf Positivity:}
The functional should be
of positive type, meaning that $\langle A^* A \rangle_\omega \ge 0$ for
each $A \in \A(\M)$.
Physically, $\langle A \rangle_\omega$ is
interpreted as the expectation value of the observable $A$ in $\omega$.

\medskip
\noindent
\paragraph{S2) \bf OPE:}
The operator product expansion holds as
an asymptotic relation. By this we mean more precisely the following.
Let $\phi^{(i_1)}, \dots, \phi^{(i_n)}$ be
any collection of fields, let $\delta >0$ be arbitrary but fixed, and
let $\T$ be any merger tree as described in the associativity condition.
Let $\sim_{\delta,\T}$ be the associated asymptotic equality
relation between distributions of $n$ spacetime points that are
defined in a neighborhood of the diagonal which was defined in
the associativity condition [see eq.~(\ref{36}]. Then we require that
\ben\label{OPEs}
\left\langle
\phi^{(i_1)}(x_1) \cdots \phi^{(i_n)}(x_n) \right\rangle_\omega
\,\, \sim_{\delta, \T} \,\,
\sum_{j}
C^{(i_1) \dots (i_n)}_{(j)}(x_1, \dots, x_n; y) \, \left\langle
\phi^{(j)}(y) \right\rangle_\omega \,\,\, ,
\een
where the sum is carried out over all $j$ such that
$d(j) \le \Delta$, where
$\Delta$ is a number depending upon the tree $\T$, and the
specified accuracy, $\delta$.

\medskip
\noindent
\paragraph{S3) \bf Spectrum condition:}
We have
\ben
\WF\Big( \langle \phi^{(i_1)}(x_1)
\dots \phi^{(i_n)}(x_n) \rangle_\omega \Big) \subset \Gamma_n(\M) \, ,
\een
where the set $\Gamma_n(\M) \subset T^*M^n \setminus \{0\}$ was defined above in eq.~\eqref{gamtdef}.

\vspace{1cm}

As part of our definition of a quantum
field theory, we make the final requirement that there is at least on state,
i.e.,
\ben
\S(\M) \neq \emptyset \quad \text{for all $\M$.}
\een
If the state space were empty, then this is a sign that the OPE is
inconsistent, and does not define a physically acceptable quantum field theory.

\medskip

\noindent
{\bf Remarks: (1)} The OPE coefficients enter the construction of the algebra
$\A(\M)$ only via condition (A4). However, they provide a strong restriction on
the state space $\S(\M)$ via condition (S2).

\noindent
{\bf (2)} If $\omega \in \S(\M)$ and $A \in \A(\M)$, then
$\omega(A^* \cdot A)$ is a positive linear functional
on $\A(\M)$ [i.e., satisfying (S1)] which can also be shown to satisfy (S2).
This functional can be identified with a
vector state in the Hilbert space representation of
$\A(\M)$ obtained by applying the GNS construction to $\omega$, and
is therefore in the domain of all smeared field operators.
It is natural to expect that, in a reasonable quantum field theory,
the OPE [i.e. (S3)] should hold in such states as well, but
this does not appear to follow straightforwardly within our axiomatic setting.

\noindent
{\bf (3)} There are some apparent redundancies in our assumptions in
that commutativity/anti-commutativity conditions have been imposed
separately on the OPE coefficients and the algebra (see conditions
(C4) and (A5)), and microlocal spectrum conditions have been imposed
separately on the OPE coefficients and the states (see conditions (C8)
and (S3)). It is possible that our assumptions could be reformulated
in such a way as to eliminate these redundancies, e.g., it is possible that
condition (A5) might follow from condition (C4) with perhaps somewhat stronger
assumptions about states.
However, we shall not pursue these possibilities here.

\medskip

The construction of the pair $\{\A(\M), \S(\M)\}$ obviously depends only
upon the data entering that construction, namely the set of all
operator product coefficients $\C(\M)$, as well as the assignments $i \mapsto
V(i)$ and $i \mapsto F(i)$ of the index set enumerating the fields
with tensor/spinor character, and with Bose/Fermi character. Thus,
any transformation on field space preserving the OPE and the
Bose/Fermi character will evidently give rise to a corresponding isomorphism
between the algebras, and a corresponding map between the state
spaces. We now give a more formal statement of this obvious fact, and
then point out some applications.

Let $L: \I \to \I, i \mapsto i' = Li$ be a map with the following
properties. (a) $L\myid = \myid$ (b) $L$ preserves the
$\star$-operation on $\I$, (c) $L$ preserves the assignment of
Bose/Fermi character, i.e., $F(i') = F(i)$. Furthermore, assume that
there is an embedding map $\psi: M \to M'$ (not necessarily an
isometry at this stage), and for each
index $i$, there is a bundle map $\psi_{(i)}$ characterized by the
following commutative diagram (where $i' = Li$)
\ben
\begin{CD}
V(i)   @>{\psi_{(i)}}>>   V(i')\\
@V{\pi_M}VV      @VV{\pi_{M'}}V \\
M    @>{\psi}>> M'
\end{CD}
\een
where $\pi_M$ respectively $\pi_{M'}$ are the bundle projections
associated with the vector bundles $V(i)$ and $V(i')$ over $M$
respectively $M'$ that characterize the spinor/tensor character of
the field labelled by $(i)$.
Recalling that $V(i^\star)$ is required to be equal to the
hermitian conjugate bundle $\overline V(i)$, and denoting by
$conj$ the operation of conjugation mapping between these bundles, we
require as a consistency condition that
\ben\label{cons0}
{\rm conj}_{\M'} \circ \psi_{(i)} = \psi_{(i)^\star} \circ {\rm conj}_\M \, ,
\een
for all indices $i \in \I$.
We say that a collection of OPE coefficients $\C(\M)$ on $\M = (M,g,T,e)$
and a collection $\C'(\M')$ on $\M' = (M',g',T',e')$ are equivalent if

\ben\label{cons1}
(\psi_{(i_1)}^* \times \cdots \times \psi_{(i_n)}^* \times \psi_{(j) *}^{-1}) \,
C^\prime{}^{(Li_1) \dots (Li_n)}_{(Lj)}[\M'] \sim C^{(i_1) \dots (i_n)}_{(j)}[\M]
\, ,
\een
see eq.~\eqref{simdef}.
By simply going through the definitions of the algebra $\A(\M)$
and the state space $\S(\M)$ it is then clear that the following
(almost trivial) lemma holds.

\begin{lemma}\label{lemma1}
Under the consistency conditions~\eqref{cons0} and~\eqref{cons1},
and assuming that $\psi$ preserves all background structure (i.e., $\psi$
is an isometric embedding preserving the
causality relations, orientations, and spin structures)
the map $\alpha_L: \A(\M) \mapsto \A(\M')$
\ben
\alpha_L: \phi^{(i)}_\M(f) \mapsto \phi^{(i')}_{\M'}(\psi_{(i) *} f),
\quad i' = Li
\een
defines a linear *-homomorphism. The dual map $\alpha_L^{\rm v}$ between the
corresponding state spaces defines a map $\S(\M') \to \S(\M)$.
\end{lemma}

Another way of stating this result is to view
the maps $L$ as described above as {\em morphisms} in
the category whose objects are the OPE-coefficient systems $\C(\M)$.
The above lemma then says that the constructions of $\S(\M)$
and of $\A(\M)$ from $\C(\M)$ are functorial in
nature.

We now discuss some applications of the lemma:

\medskip
\noindent
\paragraph{Application 1:}
Consider the case where $L=id$, $\psi: M \to M'$ is an isometric embedding
preserving orientations and any other background
structure, and
$\psi_{(i)}$ is the natural bundle map associated with $\psi$.
Then~\eqref{cons0} obviously holds,
while~\eqref{cons1} holds because of the locality and covariance
property of the OPE coefficients. The map $\alpha_L$, whose existence
is guaranteed by the lemma, then corresponds to
the map $\chi_\rho$ discussed in the
introduction. In particular, the assignment
\ben
\text{Background Structures $\to$ Algebras} \, , \quad \M \to \A(\M)
\een
is functorial, in the sense that if $\rho$ is an arrow in the
category of background structures---i.e., an orientation and
causality preserving isometric embedding from one spacetime
into another---then $\chi_\rho$ is the corresponding arrow in the
category of *-algebras---i.e., an injective *-homomorphism.
Functoriality means that the assignment
\ben
\text{Isometric Embeddings $\to$ Algebra Homomorphisms} \, , \quad
\rho \to \chi_\rho
\een
respects composition of arrows in the respective categories.
Thus, in the terminology of the introduction, the assumptions of our framework
define a local, covariant quantum field theory, and all fields
$\phi^{(i)}$ are local, covariant quantum fields. Furthermore,
$\A: \M \to \A(\M)$ is a functor in the sense of~\cite{bfv}.

\medskip
\noindent
\paragraph{Application 2:}
As the second application, consider a purely ``internal'' symmetry,
$L$, i.e., consider the case that $\M=\M'$, $\psi=id$. Then $\alpha_L$
acts upon $\A(\M)$ as a *-autmorphism, i.e., an internal symmetry.
More generally, there could be an entire group $G$ of such $L$
with the corresponding maps $\psi_{(i)}$ satisfying the composition law of the
group, i.e., if we have $L, L'$, then the map associated with
$L''=L \circ L'$ and $\psi''= \psi \circ \psi'$ is given by
$\psi_{(i)}''=\psi_{(i)} \circ \psi_{(i)}'$.
In this case, we get an action of $G$ on $\A(\M)$ by $*$-automorphisms
$\alpha_L$ satisfying the composition law $\alpha_L \circ \alpha_{L'}
= \alpha_{L''}$.

\medskip
\medskip
Another simple consequence of our axioms is
the following lemma. As above in (C6),
let $X^a$ be a vector field on $M$ locally defined near
$y$ such that $\nabla_a X^b  = -\delta_a{}^b$ at $y$.
Let $\Phi_t$ be the flow of this field\footnote{
It follows that we can write
$
\Phi_{\log \lambda}(x) = y + \lambda^{-1}(x-y)
$
in a suitable coordinate system covering $y$.}, which
scales points relative to $y$ by a factor of $e^{-t}$.
If $f$ is a compactly supported test section in $V(i)$,
we set $f_\lambda = \lambda^{-D} \, \Phi^*_{\log \lambda} f$
for $\lambda > 0$.

\begin{lemma}\label{scalelemma}
For $i \neq \myid$, there exists an $\M$, $\delta>0$, and test section $f$
such that
\ben
\lim_{\lambda \to 0} \lambda^{2\d(i)-\delta}
\int_{M \times M} C^{(i)(i^\star)}_{(\myid)}(x_1, x_2,y)
f_\lambda(x_1)
\bar f_\lambda(x_2) \,
d\mu_1 d\mu_2 = + \infty \, .
\een

\end{lemma}

{\em Proof:} Let $\omega$ be an arbitrary state. Then we have
$\langle \phi^{(i)}(f_\lambda) \phi^{(i^\star)}(\bar f_\lambda)
\rangle_\omega \ge 0$, from the star axiom (C3), and the positivity of
any state.
Let $\M$ be such that the scaling degree of
$C^{(i)(i^\star)}_{(\myid)}$ equals $2\d(i)$. By
the scaling degree and asymptotic positivity axioms (C5) and (C6),
for sufficiently small
$\delta>0$, the quantity $2\d(i) - \delta$ is bigger than
the scaling degree of
$C^{(i)(i^\star)}_{(j)}$ for any $j \neq \myid$.
Hence using eq.~\eqref{OPEs} and $\langle \myid \rangle_\omega = 1$, we have
\ben
\lim_{\lambda \to 0} \lambda^{2\d(i)-\delta} \left\{
\Bigg\langle \phi^{(i)}(f_\lambda) \phi^{(i^\star)}(\bar f_\lambda)
\Bigg\rangle_\omega
- C^{(i)(i^\star)}_{(\myid)}(f_\lambda, \bar f_\lambda)
\right\} = 0 \, .
\een
By axiom (C6), we can choose $f$ so that the second term tends to $\infty$
in absolute value as $\lambda \rightarrow 0$. However, the first term is always
non-negative. Therefore,
$C^{(i)(i^\star)}_{(\myid)}(f_{\lambda}, \bar f_{\lambda}) \rightarrow +\infty$.
\qed


\section{Normal (anti-) commutation relations}\label{section4}

In our axioms, we assumed that every field $\phi^{(i)}$ was either a
Bose or Fermi field, i.e., that there was a consistent assignment
$i \mapsto F(i) \in {\mathbb Z}_2$ such that~\eqref{causal} holds.
A priori, there is no relation between $F(i)$ and the Bose/Fermi
character of the corresponding hermitian conjugate field, i.e.,
$F(i^\star)$. We will now prove that, in fact, $F(i) = F(i^\star)$
as a consequence of our axioms.

\begin{thm}
We have ``normal (anti-)commutation relations,'' in the sense that
\ben
F(i) = F(i^\star)
\een
for all $i \in \I$.
\end{thm}

{\it Proof:} Let $f$ and $h$ be compactly supported test sections
with support in a convex normal neighborhood of a point $y \in M$.
Let $\omega$ be a quantum state, i.e., a positive normalized
linear functional $\A(\M) \to \mc$. Using~\eqref{starop}, we see that
positivity immediately implies that
\ben
\left\langle \phi^{(i)}(f) \phi^{(i)}(h) \phi^{(i^\star)}(\bar h)
\phi^{(i^\star)}(\bar f) \right\rangle_\omega \ge 0 \, .
\een
Assume now that the supports of
$f,h$ are spacelike separated. Then, using the (anti-~)commutation
relations eq.~\eqref{comm}, it
follows that
\ben\label{expect}
p \,
\left\langle \phi^{(i)}(f) \phi^{(i^\star)}(\bar f) \phi^{(i^\star)}(\bar h)
\phi^{(i)}(h) \right\rangle_\omega \ge 0 \, ,
\een
where $p = (-1)^{F(i) F(i^\star) + F(i^\star)^2}$.  Clearly, if we
could show that the expectation value in this expression were
positive for some test sections, $f,h$, in some
spacetime, then it would follow that $p = +1$, i.e.
\ben
F(i) F(i^\star) + F(i^\star)^2 = 0 \quad \text{mod $2$},
\een
and by reversing the roles of $i$ and $i^\star$, it would also follow that
\ben
F(i) F(i^\star) + F(i)^2 = 0 \quad \text{mod $2$},
\een
from which the statement $F(i)=F(i^\star)$ modulo $2$ would follow.
Clearly, it suffices to
show that the expectation value is asymptotically positive
for a 1-parameter family of
test sections $f_\lambda, h_\lambda$ whose
supports are scaled towards $y \in M$ as $\lambda \to 0$.

To show this, we consider the particular merger tree $\T$ of figure~1,
and the corresponding associativity condition. This tree corresponds
to the scaling map $\Psi_\T(\epsilon): \vec x(1) \to \vec x(\epsilon)$, with
\bena
x_1(\epsilon) &=& {\rm Exp}_y \Big( \epsilon v_1 + \epsilon^2 v_3 \Big) \non\\
x_2(\epsilon) &=& {\rm Exp}_y \Big( \epsilon v_1 + \epsilon^2 v_4 \Big) \non\\
x_3(\epsilon) &=& {\rm Exp}_y \Big( \epsilon v_2 + \epsilon^2 v_5 \Big) \non\\
x_4(\epsilon) &=& {\rm Exp}_y \Big( \epsilon v_2 + \epsilon^2 v_6 \Big) \non\\
x_5(\epsilon) &=& {\rm Exp}_y \Big( \epsilon v_1 \Big) \non\\
x_6(\epsilon) &=& {\rm Exp}_y \Big( \epsilon v_2 \Big) \non\\
x_7(\epsilon) &=& y \, .
\eena
The corresponding associativity condition together with (S2) yields
\begin{multline}
\lim_{\epsilon \to 0} \epsilon^{-\delta} \Bigg(
\Bigg\langle
\phi^{(i)}(x_1(\epsilon))
\phi^{(i^\star)}(x_2(\epsilon))
\phi^{(i^\star)}(x_3(\epsilon))
\phi^{(i)}(x_4(\epsilon))
\Bigg\rangle_\omega \\
-\sum_{j_1, j_2, j_3}
C^{(i)(i^\star)}_{(j_1)}
(x_1(\epsilon), x_2(\epsilon); x_5(\epsilon))
\, C^{(i^\star)(i)}_{(j_2)}
(x_3(\epsilon), x_4(\epsilon); x_6(\epsilon)) \\
\times
C^{(j_1)(j_2)}_{(j_3)}
(x_5(\epsilon), x_6(\epsilon); y) \,
\Bigg\langle \phi^{(j_3)}(y) \Bigg\rangle_\omega \Bigg) = 0 \, .
\end{multline}
This is to be understood in the sense of distributions in
$v_1, \dots, v_6$. The sums go over all indices with $\d(j_k) \le \Delta$, where
$\Delta$ depends on $\delta$. We now use axioms (C5) and (C6)
to analyze the scaling of the
individual terms under the sum. It follows that
\begin{equation}
\lim_{\epsilon \to 0} \epsilon^\alpha \,
C^{(i)(i^\star)}_{(j_1)} (x_1(\epsilon), x_2(\epsilon); x_5(\epsilon)) \,
C^{(i^\star)(i)}_{(j_2)} (x_3(\epsilon), x_4(\epsilon); x_6(\epsilon)) \,
C^{(j_1)(j_2)}_{(j_3)} (x_5(\epsilon), x_6(\epsilon); y)  = 0
\end{equation}
if $\alpha > 8 \d(i) - \d(j_1) - \d(j_2) - \d(j_3)$
in the sense of distributions. Thus, the term under the
sum with the potentially most singular behavior as $\epsilon \to 0$
is the one where $\d(j_1)=\d(j_2)=\d(j_3)$ is
minimal, i.e. equal to 0, by axiom (C6). If these dimensions
vanish, then by axiom (C6), we have $j_k = \myid$.

Because $\langle \myid \rangle_\omega = 1$, and because
the OPE coefficients involving only identity operators are equal to 1
by the identity axiom, we have
\begin{multline}
\lim_{\epsilon \to 0} \epsilon^{8\d(i)+\delta}
\Bigg(
\Bigg\langle
\phi^{(i)}(x_1(\epsilon))
\phi^{(i^\star)}(x_2(\epsilon))
\phi^{(i^\star)}(x_3(\epsilon))
\phi^{(i)}(x_4(\epsilon))
\Bigg\rangle_\omega \\
-
C^{(i)(i^\star)}_{(\myid)}
(x_1(\epsilon), x_2(\epsilon), x_5(\epsilon))
\, C^{(i^\star)(i)}_{(\myid)}
(x_3(\epsilon), x_4(\epsilon), x_6(\epsilon)) \Bigg) = 0 \, ,
\end{multline}
for some $\delta>0$.
We now integrate this expression against the test section
$f(v_3)\bar f(v_4) \bar h(v_5) h(v_6)$, where $f,h$ are of compact
support, and change integration
variables\footnote{
We should also integrate against a test function in $v_1, v_2$.
But the result~\eqref{rrr} is already smooth in these variables,
so we can omit this smearing.
}.
Then we get for the terms under the limit sign
\ben\label{rrr}
\Bigg\langle
\phi^{(i)}(f_{\epsilon})
\phi^{(i^\star)}(\bar f_{\epsilon})
\phi^{(i^\star)}(\bar h_{\epsilon})
\phi^{(i)}( h_{\epsilon})
\Bigg\rangle_\omega -
C^{(i)(i^\star)}_{(\myid)}(f_{\epsilon}, \bar f_{\epsilon})
C^{(i^\star)(i)}_{(\myid)}(\bar h_{\epsilon}, h_{\epsilon})
\een
Here, we have defined $f_\epsilon(x) = \epsilon^{-2D} f \circ \alpha_1(\epsilon, x)$ and
$h_\epsilon(x) = \epsilon^{-2D} h \circ \alpha_2(\epsilon, x)$ with $\alpha_i(\epsilon, \,.\,):M \to \mr^D$
are the maps that are defined in a sufficiently small neighborhood
of $y$ by
\bena\label{choice}
\alpha_1(\epsilon,x) &=& \epsilon^{-1} v_1 + \epsilon^{-2} {\rm Exp}_y^{-1} (x)\\
\alpha_2(\epsilon,x) &=& \epsilon^{-1} v_2 + \epsilon^{-2} {\rm Exp}_y^{-1} (x)\, .
\eena
Finally, we use the lemma~\ref{scalelemma} to conclude that there exist $f,h$ such that both
$C^{(i)(i^\star)}_{(\myid)}(f_\epsilon, \bar f_\epsilon) \to +\infty$ and
$C^{(i)(i^\star)}_{(\myid)}(h_\epsilon, \bar h_\epsilon) \to +\infty$ for
some spacetime and some subsequence of $\epsilon \to 0$. In view of eq.~\eqref{rrr}, it follows that
\ben
\lim_{\epsilon \to 0}
\left\langle
\phi^{(i)}(f_{\epsilon})
\phi^{(i^\star)}(\bar f_{\epsilon})
\phi^{(i^\star)}(\bar h_{\epsilon})
\phi^{(i)}( h_{\epsilon})
\right\rangle_\omega = + \infty \, ,
\een
so the expectation value~\eqref{expect} is positive
for the choice of test sections $f,h$ given by $f_\epsilon,
h_\epsilon$, see eq.~\eqref{choice}, for
sufficiently small $\epsilon$. These test sections will have
spacelike separated support as long as ${\rm Exp}_y v_1$ and ${\rm Exp}_y v_2$ are spacelike,
which we may assume to be the case. \qed

\section{The spin-statistics theorem and the PCT-theorem}\label{section5}

In this section, we prove that appropriate versions of the spin-statistics
theorem and the PCT theorem hold in curved spacetime under our axiom scheme.
We explicitly discuss the case when the spacetime dimension
is even, $D=2m$ and discuss the case of odd dimensions briefly
in remark (2) below the PCT theorem.
The key ingredient in both proofs is a relation, proven in
in~\cite{hpct}, between
the OPE coefficients $\C(\M)$ on the background structure $\M
= (M,g,T,e)$,
and the OPE coefficients $\C({\overline{\M}})$ on the background structure
\ben
\overline \M = (M,g,-T,e)
\een
consisting of the same manifold $M$,
the same metric $g$, the same orientation $e$, but
the {\em opposite} time orientation $T$. For even spacetime
dimensions $D=2m$, this relation is\footnote{Note that the
``bar'' symbol is referring to the $PT$-reversed background
structure in the term on the left side, while it means
hermitian conjugation on the right side.}:
\ben\label{consist}
 C^{(i_1) \dots (i_n)}_{(j)}[\overline \M]
\sim
\overline{C^{(i_1^\star) \dots
  (i_n^\star)}_{(j^\star)}[\M]} \cdot
\begin{cases}
\i^{-F(j)}(-1)^{-U(j)} \prod_{k=1}^n \i^{F(i_k)}(-1)^{U(i_k)} &
\text{$m$ even,} \\
\i^{-F(j)+U(j)-P(j)} \,\, \prod_{k=1}^n \i^{F(i_k)-U(i_k)+P(i_k)} &
\text{$m$ odd,}
\end{cases}
\een
where $\i = \sqrt{-1}$. Here, we recall that with each
quantum field $\phi^{(i)}$ there is associated
a bundle $V(i)$ over $M$ corresponding to the
tensor or spinor character of the quantum field.
In even spacetime dimensions $D=2m$, such a bundle
$V(i)$ is a tensor product
\ben\label{Videf}
V(i) = S_-^{\otimes U(i)} \otimes S_+^{\otimes P(i)},
\een
where the first factor corresponds to the $U(i)$ ``unprimed-'' and the second
to the $P(i)$ ``primed'' spinor indices. More precisely, the bundles $S_\pm$
are defined as the $\pm 1$ eigenspaces of the chirality operator\footnote{
Here, the orientation $D$-form is normalized so that
$g^{a_1 b_1} \dots g^{a_D b_D} e_{a_1 \dots a_D} e_{b_1 \dots b_D} = -D!$.
}
\ben\label{chirality}
\Gamma = \frac{1}{D!} (-1)^{(m-1)(2m-1)/2} \, e^{a_1 a_2 \dots a_D} \gamma_{a_1} \gamma_{a_2}
\cdots \gamma_{a_D} \,, \quad\quad \Gamma^2 = id_S \,\,,
\een
acting on a $2^m$-dimensional complex vector bundle $S$ over $M$ of ``Dirac spinors''.
This bundle $S$ corresponds to a fundamental representation of the Clifford
algebra (in the tangent bundle) generated by the curved space gamma-matrices $\gamma_a$.
There exists a linear isomorphism $c: S \to \overline S$, where $\overline S$ is
the conjugate bundle of anti-linear maps $S^{\rm v} \to \mc$. Owing to the relation $\overline \Gamma = (-1)^{m-1}
\, c \, \Gamma \, c^{-1}$, it follows that for even $m$, the bundles
$S_\pm$ and $\overline S_\mp$ are isomorphic via $c$, while for odd $m$ the bundles $S_\pm$
and $\overline S_\pm$ are isomorphic, and we will hence always identify these bundles.
Thus, for even $m$ the roles of primed and unprimed spinor indices [i.e., respective tensor factors
in eq.~\eqref{Videf}] are exchanged when passing to the hermitian adjoint
$\phi^{(i^\star)}$ of a quantum field $\phi^{(i)}$, while for odd $m$
the roles are not exchanged.

We also note that the coefficients on the
right side of eq.~\eqref{consist} are sections in the (tensor product of the) spin
bundles $V_\M(i)$ referring to the time function $T$ associated
with $\M$, while the coefficients on the left side are sections
in the spinor bundles $V_{\overline \M}(i^\star)$ defined via the
opposite time orientation $-T$ associated with $\overline \M$.
As explained in~\cite{hpct}, there is a natural identification
map between these bundles, and this identification map is understood
in~\eqref{consist}.

\medskip

The proof of~\eqref{consist} makes use of the microlocal, analytical, and causal
properties of the OPE coefficients and proceeds via analytic
continuation~\cite{hpct}.
Since it is the main input in the
proofs of both the spin-statistics theorem
and the PCT theorem, we now outline, following~\cite{hpct}, how~\eqref{consist}
is proven
within our axiomatic setting. We first consider the case where $g$
is analytic.
Let $y \in M$
and introduce Riemannian normal coordinates
$x=(x^0, \dots, x^{D-1}) \in \mr^D$
about $y$. In this neighborhood of $y$, consider the 1-parameter
family of metrics $g^{(s)}$ for all $|s| \leq 1$
defined by
\ben
g^{(s)} = g_{\mu\nu}(sx) \, dx^\mu dx^\nu \, .
\een
Note that this family, in effect, interpolates between the given
metric $g=g^{(1)}$ and the
flat Minkowski metric $\eta=g^{(0)}$. We can expand $g^{(s)}$ in a power series
in $s$ about $s=0$, which takes the form
\ben
g^{(s)}_{\mu\nu} = \eta_{\mu\nu}^{} + \sum_{n=2}^\infty s^n
p_{\mu\nu \beta_1 \dots \beta_{n-2}}(y) \, x^{\beta_1} \dots x^{\beta_{n-2}}
\een
where each $p$ is a curvature polynomial $p(y)=p[
R_{\mu\nu\sigma\rho}(y), \dots, \nabla_{(\alpha_1} \cdots
\nabla_{\alpha_{(n-2)})} R_{\mu\nu\sigma\rho}(y)]$. It can then be
shown, using axiom (C1), that each OPE coefficient
has an asymptotic expansion of the form
\bena
C^{(i_1) \dots (i_n)}_{(j)}(x_1, \dots, x_n, y) = \sum_{k=0}^\infty q_k(y) \cdot (W_k)^{(i_1) \dots (i_n)}_{(j)}(x_1, \dots,
x_n) \, ,
\eena
where $q_k = (q_k)_{\mu_1 \dots \mu_k}$ is a curvature polynomial of the same general
form as the $p$, and
where $W_k = (W_k)^{\mu_1 \dots \mu_k}$ are distributions defined
on a neighborhood of $0$ in $(\mr^D)^n$,
valued in the tensor product of $(\mr^D)^{\otimes k}$ with the spinor representation corresponding
to the index structure of the quantum fields in the operator product
considered. They transform covariantly under the connected component
${\rm Spin}_\mr(D-1,1)_0$ of the spin group of $D$-dimensional
Minkowski space.

Consider now the
map $\rho$ defined in a suitable convex normal neighborhood,
${\mathcal O}$, of $y$ by
$(x^0, \dots, x^{D-1}) \mapsto (-x^0, \dots, -x^{D-1})$. In Minkowski spacetime,
this map would define an isometry which preserves
spacetime orientation but reverses time orientation. In a general curved
spacetime, this map does not define an isometry. Nevertheless, we may
view $\rho$ as a map
\ben
\rho : ({\mathcal O}, g^{(s)}, e, T) \to ({\mathcal O}, g^{(-s)}, e, -T) \, .
\een
Viewed in this manner, it is easily seen that $\rho$
preserves all background structure, i.e., it is
a causality preserving isometry
that preserves orientations. Consequently, by the covariance axiom
(C1), the relation~\eqref{consist} is equivalent to a corresponding
relation between the
OPE-coefficients on the spacetimes $(M,g^{(s)},e,T)$ and
$(M,g^{(-s)},e,T)$, i.e., spacetimes with different metrics but
the same orientation and time
orientation. If one now differentiates this relation $m$-times with respect to
$s$ and puts $s=0$ afterwards, then one can prove that~\eqref{consist}
is equivalent to the relation
\begin{eqnarray}
\label{WCPT}
&&W_k(x_1, \dots, x_{n}) = \\
&&(-1)^k \, \pi^* W_k(-x_n, \dots, -x_1) \cdot
\begin{cases}
\i^{-F(j)}(-1)^{-U(j)} \prod_{k=1}^n \i^{F(i_k)}(-1)^{U(i_k)} &
\text{$m$ even,} \\
\i^{-F(j)-U(j)+P(j)} \,\, \prod_{k=1}^n \i^{F(i_k)-U(i_k)+P(i_k)} &
\text{$m$ odd,}
\end{cases}
\non
\end{eqnarray}
for all $k=0,1,2,\dots$. Here, $\pi$ is the permutation~\eqref{permutationpi},
which acts by permuting the implicit spinor/indices associated with the spacetime
points $x_i$.

We have thus reduced the proof of~\eqref{consist}
to the proof of a statement about Minkowski
distributions $W_k$ that transform covariantly under ${\rm Spin}_\mr(D-1,1)_0$.
To prove it, one next shows that $W_k$ can be analytically
continued, and that the analytic continuation transforms covariantly under
connected component of the identity in
the complexified spin group ${\rm Spin}_\mc(D-1,1)_0$. For this, one first
proves, using the microlocal condition on the OPE-coefficients,
that, near $0$, the analytic wave front set~\cite{hormander}, $\WF_A$,
of $W_k$ satisfies
\begin{equation}\label{wavefrontset}
\WF_A(W_k) \subset K.
\end{equation}
Here, $K$ is a conic set defined
in terms of the Minkowskian metric $\eta$ and orientation $e,T$,
by
\begin{eqnarray}
\label{uawfs}
K &=&
\Big\{(y_1, k_1; \dots; y_n, k_n)
\in T^*(\times^n B_r) \setminus \{0\} \,\, \Big| \,\,
\text{$\exists p_{ij} \in \bar V_+$, $n \ge j > i \ge 1$:} \nonumber\\
&& \text{
$k_i = \sum_{j:j>i} p_{ij} - \sum_{j:j<i} p_{ji}$ for all $i$}
\Big\},
\end{eqnarray}
where $\bar V^+$ is the closure of the forward light cone $V^+$ in
Minkowski space (defined with respect to the time orientation $T$),
\begin{equation}
V^\pm = \{ k \in \mr^D \mid \eta_{\mu\nu} k^\mu k^\nu > 0, \quad
k^\mu \nabla_\mu
T > 0\}.
\end{equation}
The relation~\eqref{wavefrontset} is important because
a theorem of~\cite{hormander} now guarantees that $W_k$ is the
distributional boundary value
\ben\label{bvformula}
W_k(x_1, \dots, x_n) =
\bv_{(y_1, \dots, y_n) \in K^{\rm v} \to 0} W_k(x_1+\i y_1, \dots, x_n+\i y_n)
\een
of
a holomorphic function $W_k(z_1, \dots, z_n)$ that is defined in
the ``half-space''
\ben\label{analdomain}
W_k: B_r(0)^n + \i K^{\rm v} \to \mc, \quad \text{some $r>0$}
\een
where $B_r(0)$ is a ball of radius $r$ in $\mr^D$,
where $K^{\rm v}$ is the ``dual cone'' of all covectors
$(y_1, \dots, y_n) \in {(\mr^D)^n}$ with the property that
$\sum k_i \cdot y_i>0$ for all $(k_1, \dots, k_n) \in K$.
Using the ``edge of the wedge-theorem''~\cite{sw}, one proves
that the holomorphic function $W_k(z_1, \dots, z_n)$
transforms
covariantly under the
spin group ${\rm Spin}_\mr(D-1,1)_0$. As explained in more detail
in~\cite{hpct}, one can use this in turn
to prove the desired relation~\eqref{WCPT}:

For $D=2m$ and $m$ even, we consider
the chirality element $\Gamma$ in eq.~\eqref{chirality} in flat space, which
is an element of the connected component of the identity of the complexified spin group
${\rm Spin}_\mc(D-1,1)_0$. It corresponds to the
reflection element $\rho: \mr^D \to \mr^D, x \mapsto -x$ of the
complexified Lorentz group $SO(D-1,1; \mc)$ under the
standard covering homomorphism between these groups.
This is an immediate consequence of the relation
$\Gamma \gamma_a \Gamma^{-1} = -\gamma_a$. Using
the method of analytic continuation in overlapping patches, it can be shown
that $W_k$ may be continued to a single valued analytic function
on an extension of the domain in eq.~\eqref{analdomain}, and it can be shown
that this continuation transforms covariantly under $\Gamma$. As explained above,
$\Gamma$ acts as $+id_{S_+}$ on each tensor factor
corresponding to a primed spinor index, and as $-id_{S_-}$
on each tensor factor associated with an unprimed spinor index.
Therefore, if we apply the transformation law of $W_k$ under
the element $\Gamma$,
then we obtain relation~\eqref{WCPT} for complex spacetime
arguments, except that
the order of the complex spacetime arguments $z_1, \dots, z_n$
is reversed, and except for the factors relating to the
Bose-Fermi character of the fields involved. In order to
be able to take the limit ${\rm Im} \, z_i \to 0$ from within
$K^{\rm v}$, we must pass to so-called ``Jost points'' $(z_1, \dots, z_n)$
in the extended domain
of holomorphicity. For such points, the (anti-)commutativity may
be used, effectively allowing to permute the spacetime arguments
in $W_k$ in such a way that one can take the limit to real
points from within $K^{\rm v}$ as required in eq.~\eqref{bvformula} afterwards. When permuting the
arguments, we pick up the factors related to the Bose/Fermi character
of the fields.

For $D=2m$ and $m$ odd, we consider instead the element $\i \Gamma$ of
the connected component of the identity of the
complexified spin group ${\rm Spin}_\mc(D-1,1)_0$.
This element again covers the reflection $\rho(x)=-x$ on
$D$-dimensional Minkowski space. It acts as $+\i \, id_{S_+}$ on each tensor factor
corresponding to a primed spinor index, and as $-\i \, id_{S_-}$
on each tensor factor associated with an unprimed spinor index.
Again it can be shown that $W_k$ may be continued analytically
to a domain extending that in eq.~\eqref{analdomain}, and that
it transforms covariantly under $\i \Gamma$ on the extended domain.
The additional factor of $\i$ gives rise to the different factors in
eq.~\eqref{WCPT} compared to the case when $m$ is even.
The rest of the argument is identical to that case.

This proves the PCT-relation~\eqref{consist} for analytic spacetimes,
and even dimensions $D=2m$.
The validity of the corresponding relation for smooth spacetimes
then follows from the smoothness of the OPE-coefficients under
smooth variations of the metric, since any smooth metric can be
viewed as the limiting member of a smooth 1-parameter family of
metrics $g^{(\lambda)}$ that are analytic for $\lambda > 0$ and
smooth for $\lambda = 0$. The differences in the statement and proof of the
PCT-relation~\eqref{consist} for odd spacetime dimension $D$
are described in remark (2) below, following the proof of the PCT-Theorem.

\bigskip

We now are ready to state and prove the spin-statistics theorem
within our framework. The statement and proof of this theorem closely
parallel the Minkowski spacetime version:

\begin{thm} ({\em Spin-Statistics Theorem})
If our axioms hold, then the
spin statistics relation
\ben
F(i) = U(i) + P(i) \quad mod \, 2 \, ,
\een
also holds,
i.e. fields with integer spin (= on half the number of primed + unprimed
spinor indices) have Bose statistics, while fields of half integer spin
have Fermi statistics.
\end{thm}

\noindent
{\it Proof:}
Let $i \in \I$, and, as above, we restrict consideration to the
even dimensional case $D=2m$.
Consider the PCT-relation~\eqref{consist} for the
OPE-coefficient $C^{(i)(i^\star)}_{(\myid)}$. This condition can be
written as
\ben
C^{(i)(i^\star)}_{(\myid)}(x_1, x_2, y)_\M
\sim
C^{(i^\star)(i)}_{(\myid)}(x_2,
x_1; y)_{\overline \M} \cdot
\begin{cases}
\i^{F(i)+F(i^\star)}
(-1)^{U(i^\star)+U(i)} & \text{$m$ even,} \\
\i^{F(i)+U(i)-P(i)+F(i^\star)+U(i^\star)-P(i^\star)} &
\text{$m$ odd,}
\end{cases}
\een
where we have used the hermitian conjutation
axiom, and
where we have used that $F(\myid)=0$ since the identity
is always a Bose field, by the identity axiom.
When $m$ is even, then
$U(i^\star) = P(i)$ because conjugation of a spinor exchanges the
number of primed and unprimed indices. Furthermore, $F(i) =
F(i^\star)$ by Theorem~1, so we obtain
\ben\label{1}
C^{(i)(i^\star)}_{(\myid)}(x_1, x_2, y)_\M
\sim (-1)^{F(i)+U(i)+P(i)} C^{(i^\star)(i)}_{(\myid)}(x_2,
x_1; y)_{\overline \M} \, .
\een
When $m$ is odd, $U(i^\star) = U(i)$ and $P(i^\star) = P(i)$,
because conjutation of a spinor does not change
the number of primed and unprimed spinor indices in that case. Using
this, we again obtain the expression eq.~\eqref{1} when $m$ is odd.

We now smear this expression with the test section $f_\lambda(x_1) \bar
f_\lambda(x_2)$, where
\ben
f_\lambda(x) = \lambda^{-D} f[y+\lambda^{-1}(x-y)] \, .
\een
We are taking here an $f$ of compact support in a sufficiently
small neighborhood of $y$ covered by some coordinate system, and
$y+\lambda^{-1}(x-y)$ is computed in this
arbitrary coordinate system. Denote by $\d(i)$ the dimension of the
field labelled by $i$. It follows from eq.~\eqref{1} that
\begin{multline}
\lambda^{2\d(i)-\delta} \int
C^{(i)(i^\star)}_{(\myid)}(x_1, x_2; y)_\M
\,
f_\lambda(x_1) \bar f_\lambda(x_2) \,
d\mu_1 d\mu_2 \\
- p
\, \lambda^{2\d(i)-\delta} \int
C^{(i^\star)(i)}_{(\myid)}(x_2, x_1; y)_{\overline \M}
\,
f_\lambda(x_1) \bar f_\lambda(x_2) \,
d\mu_1 d\mu_2 \to 0
\end{multline}
as $\lambda \to 0$, where $p=(-1)^{F(i)+U(i)+P(i)}$.
But the first term goes to
$+\infty$ by lemma~\ref{scalelemma}, while the
second term goes to $-p \cdot \infty$ for suitable $f$.
Thus, these terms can only cancel for small $\lambda$
if we have $p=+1$, meaning
that $F(i) + P(i) + U(i) = 0$ modulo 2. Hence
the spin-statistics relation must hold. \qed

\bigskip

Next, we state and prove the PCT theorem, the formulation of which is quite
different from the Minkowski spacetime version (see the discussion below).
Again, we restrict consideration here to $D=2m$, and describe the differences
occurring in odd dimensions in remark (2) below:

\begin{thm} ({\em PCT-Theorem})
Given a background structure $\M=(M,g,e,T)$, we write $\overline \M =
(M,g,e,-T)$. In spacetime dimension $D=2m$, define the anti-linear map
$\theta_\M^{\rm PCT}: \A(\M)\rightarrow \A(\overline \M)$
by
\ben\label{PCT}
\theta_\M^{\rm PCT}: \phi_\M^{(i)} (f) \mapsto
\phi_{\overline \M}^{(i)}(f)^* \cdot
\begin{cases}
\i^{F(i)} (-1)^{U(i)} & \text{when $m$ is even,} \\
\i^{F(i)+U(i)-P(i)}   & \text{when $m$ is odd.}
\end{cases}
\een
Then $\theta_\M^{\rm PCT}$ is an anti-linear *-isomorphism
such that the diagram
\ben
\begin{CD}
\A(\M)   @>{\theta_{\M}^{\rm PCT}}>>   \A(\overline \M)\\
@V{\chi_\rho}VV      @VV{\chi_\rho}V \\
\A(\M')    @>{\theta_{\M'}^{\rm PCT}}>> \A(\overline \M')
\end{CD}
\een
as well as the diagram
\ben
\begin{CD}
\S(\overline \M')   @>{\theta_{\M'}^{\rm PCT \, v}}>>   \S(\M')\\
@V{\chi_\rho^{\rm v}}VV      @VV{\chi_\rho^{\rm v}}V \\
\S(\overline \M)    @>{\theta_{\M}^{\rm PCT \, v}}>> \S(\M)
\end{CD}
\een
commute for every isometric, causality and orientation preserving
embedding $\rho: \M \to \M'$. Here $\chi_\rho^{\rm v}$ denotes the dual
of the linear map $\chi_\rho$, and $\theta^{\rm PCT \, v}$
denotes the dual of $\theta^{\rm PCT}$.
\end{thm}
\medskip
\noindent
{\em Proof:}
The proof of this theorem is, in essence, an application of lemma~\ref{lemma1}.
In the notation of lemma~\ref{lemma1},
we choose $\M=(M,g,e,T)$, we choose $\M' =
\overline \M$, we take $\psi=id$ and we choose
$L: i \mapsto i^\star$. For any index $i$, we define $\psi_{(i)}$
to be the composition of the
natural anti-linear bundle map from $V_\M(i)$ to $V_{\overline \M}(i^\star)$ that
is implicit in the formula~\eqref{PCT} with the multiplication map by
$\i^{F(i)} (-1)^{U(i)}$ when $m$ is even and by $\i^{F(i)+U(i)-P(i)}$ when $m$ is odd.
From this definition we then have
\ben
{\rm conj}_{\overline \M} \circ \psi_{(i)} =
\begin{cases}
(-1)^{F(i)/2+F(i^\star)/2+U(i)-U(i^\star)} \,\,\,\,\,\,
\cdot \psi_{(i^\star)} \circ {\rm conj}_\M & \text{$m$ even,}\\
\i^{F(i)+F(i^\star)-U(i)-U(i^\star)+P(i)+P(i^\star)}
\cdot \psi_{(i^\star)} \circ {\rm conj}_\M
& \text{$m$ odd,}
\end{cases}
\een
where $conj_{\M}$ is the anti-linear map that sends a spinor to the
hermitian conjugate spinor on $M$.
Now for even $m$ the number of unprimed spinor
indices associated with $V(i)$ is precisely equal to the number
of primed indices $P(i^\star)$ associated with
$V(i^\star)$, because $V(i^\star)$ is assumed to
be equal to $\overline V(i)$.
Thus, $P(i)=U(i^\star)$. Furthermore, we have $F(i)=F(i^\star)$ by Theorem~1,
and $F(i) = U(i) + P(i)$ mod $2$ by the spin-statistics theorem. Consequently,
the compatibility condition for the *-operation holds when $m$ is even, i.e.,
\ben\label{compatib}
{\rm conj}_{\overline \M} \circ \psi_{(i)} = \psi_{(i^\star)} \circ {\rm conj}_\M \, .
\een
When $m$ is odd, then $U(i) = U(i^\star)$ and $P(i) = P(i^\star)$, and
the compatibility condition again holds because of $F(i) = F(i^\star)$ and
the spin-statistics theorem. Thus, we have shown that the first
input in lemma~\ref{lemma1} holds.
The second input is the compatibility of the OPE coefficients
on $\M$ and $\M'$, eq.~\eqref{cons1}. That condition is essentially
equivalent to the relation~\eqref{consist}, except that
the latter relation also involves an
additional complex conjugation of the OPE-coefficient.
However it is immediately seen that
this will result only in the following difference in the conclusion
of lemma~\ref{lemma1}: Instead of the linear
*-homomorphism as provided by this lemma,
we now find that
the PCT-map $\theta^{\rm PCT}$
defined by eq.~\eqref{PCT} yields an anti-linear *-homomophism.
\qed

\medskip

\paragraph{Remarks:}
(1) The above formulation of the PCT-theorem was suggested in
\cite{hpct}. As noted in~\cite{Fredlisbon}
the theorem can be stated in the language of functors by saying that the
functors $\M \to \A(\M)$ and $\M \to \overline \A(\M) = \A(\overline
\M)$ are equivalent.

\noindent
(2) In odd spacetime dimensions $D=2m+1$, the chirality operator
$\Gamma$ of eq.~\eqref{chirality} is proportional to the identity in
$S$. Thus, in this case there is no decomposition $S=S_+\oplus S_-$
as in the even dimensional case, and there is consequently no difference
between ``primed'' and ``unprimed'' spinors. If we denote the
number of spinor indices of a quantum field by $N(i)$ (i.e., the
bundle associated with the label $i$ the field is $V(i) = S^{\otimes N(i)}$),
then the factors in formula~\eqref{consist} are now $\i^{-F(j)-N(j)} \prod
\i^{F(i_k)+N(i_k)}$. In the proof of this formula, one must now consider
the map $\rho: (x^0, x^1, \dots, x^{D-1}) \mapsto (-x^0, -x^1, \dots, +x^{D-1})$.
This corresponds again to a change of time orientation in Minkowski spacetime
which leaves the spacetime
orientation invariant. The rest of the proof is similar.

\medskip

We now explain the
relation of the above formulation of the PCT theorem to
to the usual PCT theorem in Minkowski spacetime (see e.g.~\cite{sw}).
Changing $T \to -T$ while keeping $e$ unchanged is
equivalent to changing parity (i.e., the spatial
orientation $s$ of a Cauchy surface
$\Sigma$ induced by $e$ and $T$ via $dT \wedge s = e$) and time
(i.e., the time function). Furthermore, the field appearing on the right
side of eq.~\eqref{PCT} is usually referred to as the ``charge conjugate
field'' to $\phi_\M^{(i)} (f)$.
Thus our formulation of the PCT
theorem asserts that the theory
is indeed invariant under simultaneous PCT-reversal in the sense
that the theory on $\M$ is ``the same'' as the theory on $\overline \M$
with the fields replaced by their charge conjugates. However, note that
our PCT theorem relates theories on the two {\it different} background
structures, $\M$ and $\overline \M$. By contrast, the usual PCT
theorem in Minkowski spacetime provides a symmetry of the theory
defined on a single background structure, namely Minkowski spacetime with a
fixed choice of orientation and time orientation. Indeed, the usual
formulation of the PCT theorem in Minkowski spacetime asserts the
existence of an
anti-unitary operator $\Theta: {\mathcal H} \to {\mathcal H}$
on the Hilbert space, ${\mathcal H}$, of
physical states such that, if $\rho$ denotes the
isometry on the, say, even-dimensional Minkowski spacetime defined by
\ben\label{rhodef}
\rho: (x^0, x^1, \dots, x^{D-1}) \mapsto (-x^0, -x^1, \dots, -x^{D-1}),
\een
and if $\phi^C$ is the
charge conjugate field associated with $\phi$ defined by eq.~\eqref{PCT},
then ${\rm Ad}_\Theta \phi(\rho(x)) \equiv \Theta \phi(\rho(x))
\Theta^{\dagger} = \phi^C(x)$.

The relationship between these formulations can be seen as follows.
Start with our formulation of the PCT theorem.
The isometry
$\rho$ maps Minkowski spacetime
$\M = (\mr^D, \eta, e,T)$ with a given choice of orientation $e$ and time
orientation $T$, to $\overline \M = (\mr^D,
\eta, e, -T)$. Thus, by ``application 1'' of lemma~\ref{lemma1},
we know that there is a *-homomorphism
$\chi_\rho: \A(\M) \to \A(\overline \M)$ mapping the quantum fields
$\phi_\M(f)$ on $\M$ to the ``same'' quantum fields
$\phi_{\overline \M}(\rho_*(f))$ on $\bar \M$. Thus, if we define
\ben\label{relation1}
{\rm Ad}_\Theta := \chi_\rho^{-1} \circ \theta^{\rm PCT}_{\M}   \, .
\een
we obtain a result that is essentially equivalent to
the usual Minkowski version (as suitably reformulated in an algebraic
setting). Conversely, if we start with the usual formulation of the PCT
theorem and if we define quantum field theory on $\A(\overline \M)$ in terms of
quantum field theory on $\M$ by means of the map $\chi_\rho$,
then we obtain a version essentially equivalent to our formulation by
setting
\ben\label{relation}
\theta^{\rm PCT}_{\M} = \chi_\rho \circ {\rm Ad}_\Theta \, .
\een

Although the above formulations are essentially equivalent in Minkowski
spacetime, in a general spacetime, there does
not exist any discrete isometry analogous to $\rho$.
Thus, in general we only have a PCT
theorem describing the relation between the theory defined on {\it
different} backgrounds $\M$ and $\overline \M$. Of course
in the case of a spacetime that admits an isometry $\rho$
mapping $(e,T)$ to $(e, -T)$
(as occurs, e.g. in Schwarzschild and deSitter spacetimes), then
a ``same background structure'' version of the PCT theorem can
be given via eq.~\eqref{relation1}.

The example of a Robertson-Walker spacetime
\ben
\label{RW}
g = -dt^2 + a(t)^2 \, d{\bf x} \cdot d{\bf x}
\een
with $a(t)$ a strictly increasing function
of $t$, may be useful for clarifying the physical meaning of our
formulation of the PCT theorem in a general curved spacetime. If we
choose the time orientation $T=t$, then the above metric describes an
expanding universe, while if we take $T=-t$, it describes a corresponding
contracting universe (with opposite choice of spatial orientation since
we keep $e$ fixed).
In essence, our formulation of the PCT theorem relates
phenomena/processes occuring in the {\it expanding} universe eq.~\eqref{RW} to
corresponding processes (involving the charge conjugate fields and also
a reversal of parity) in the corresponding {\it contracting} universe.
Since the metric eq.~\eqref{RW} has no
time reflection isometry $\rho$, there are no relations implied by the
PCT theorem between phenomena/processes occurring in the
expanding universe, eq.~\eqref{RW} with $T=t$. As a concrete illustration
of this, suppose that it were possible to give a definition of ``particle
masses'' in curved spacetime---although it is far from obvious that any such
useful notion exists. The PCT theorem would then imply that the mass
of a particle in an expanding universe
must be equal to the mass of the corresponding antiparticle in a contracting
universe. However, it would make no statement about the masses of particles
and antiparticles in the same universe\footnote{It is worth noting that
the ``third Sakharov necessary condition''
for baryogenesis in the early
universe (namely, ``interactions out of thermal equilibrium'') is based
upon the (now seen to be unjustified) assumption that particle and
antiparticle masses are equal in an expanding universe.
However, to the extent that particle and antiparticle masses might differ
in an expanding universe (even assuming that a useful
notion of ``particle mass''
can be defined) as a result of the lack of a time reflection symmetry,
it would probably not even be possible to define a notion
of ``thermal equilibrium'' as a result of the lack of a time translation
symmetry.}.

Finally, it is worth emphasizing the nature of the action of our
PCT map $\theta_{\M'}^{\rm PCT \, v}$ on states.
In Minkowski spacetime, the states of interest are normally
assumed to lie in a single
Hilbert space $\H$, and one often considers scattering states. In the
usual formulation of the PCT theorem in Minkowski spacetime, if
$\omega$ describes an incoming scattering state $|p_1, \dots, p_n;
{\rm in} \rangle \in \H$, then the corresponding state $\bar \omega$
under the PCT map may be
identified with an outgoing scattering state $|-p_1, \dots, -p_n; {\rm
out} \rangle \in \H$ in the {\em same} Hilbert space. However, in our
formulation of the PCT theorem, if $\omega$ is an in-state in $\M$,
then $\bar \omega$ is an
in-state in $\overline \M$. If the spacetime admits a time reflection isometry
$\rho$, then $\bar \omega$ may be identified with a
corresponding out-state on $\M$. However, for the Robertson-Walker
spacetime
eq.~\eqref{RW}, which does not admit a time reflection isometry,
the PCT theorem relates in-states in an expanding universe to
in-states in the corresponding contracting universe.

\section{Conclusions and outlook}

In this paper we have proposed a new axiomatic framework for quantum
field theory on curved spacetime. We demonstrated that our new
framework captures much of the same content as the Wightman axioms by
proving curved spacetime analogs of the spin-statistics theorem and
PCT theorem. In this section, we discuss some of the implications and
potential ramifications of the viewpoint suggested by this new
framework.

First, we address the issue of why we even seek an axiomatic framework
for quantum field theory in curved spacetime at all. Since gravity is
being treated classically, quantum field theory in curved spacetime
cannot be a fundamental theory of nature, i.e., it must have a limited
domain of validity. In particular, we have focused considerable
attention in this paper on the OPE of quantum fields in curved
spacetime, but the OPE is a statement about the
arbitrarily-short-distance singularity structure of products of
fields. One would not expect that quantum field theory in curved
spacetime would give an accurate description of nature at separations
smaller than, say, the Planck scale. Consequently, why should one seek
a set of mathematically consistent rules governing quantum field theory
that are rigorously applicable only
in a regime where the theory is not expected to be valid?

Our response to this question is that an exactly similar situation
arises for classical field theory. Classical field theory also is not
a fundamental theory of nature, and its description of nature makes
essential use of differentiability/smoothness properties of the
classical fields at short distance scales; one could not even write
down the partial differential equations governing the evolution of
classical fields without such short-distance-scale
assumptions. However, if quantum field theory is any guide, the
description of physical fields as smooth tensor fields is drastically
wrong at short distance scales.  Nevertheless, classical field theory
has been found to give a very accurate description of nature within
its domain of validity, and we have obtained a great deal of insight
into nature by obtaining a mathematically precise formulation of
classical field theory.  It is our belief that there exists a
mathematically consistent framework for quantum field theory in curved
spacetime, and that by obtaining and studying this framework, we will
not only get an accurate description of nature within the domain of
validity of this theory, but we will also get important insights and
clues concerning the nature of quantum gravity.

We began our quest for the mathematical framework of quantum field
theory in curved spacetime by seeking to generalize the Wightman
axioms to curved spacetime in as conservative a manner as possible. As
described much more fully in the Introduction, there are three key
ingredients of the Wightman axioms that do not generalize
straightforwardly to curved spacetime: (1) Poincare invariance; (2)
the spectrum condition; (3) existence of a Poincare invariant
state. We have seen in this paper that quantum field theory can be
generalized to curved spacetime by replacing these ingredients by the
following: (1') quantum fields are locally and covariantly defined;
(2') the microlocal spectrum condition; (3') existence of an OPE.
Although the formulation of conditions (1') and (2') differs
significantly from the formulation of conditions (1) and (2), the
basic content of these conditions is essentially the same. Indeed,
there would be no essential difference in the formulation of axiomatic
quantum field theory in Minkowski spacetime if one replaced (1) and
(2) with (1') and (2').  By contrast, as we shall elucidate further
below, the replacement of (3) by (3') leads to a radically different
viewpoint on quantum field theory.

The most important aspect of this difference is that the existence of
a ``preferred state'' no longer plays any role in the formulation of
the theory. States are inherently non-local in character, and the
replacement of (3) by (3')---along with the replacements of (1) with
(1') and (2) with (2')---yields a formulation of quantum field theory
that is entirely local in nature. In this way, the formulation of
quantum field theory becomes much more analogous to the formulation of
classical field theory. Indeed, one can view a classical field theory
as being specified
by providing the list of fields $\phi^{(i)}$ occuring in the theory
and the list of local, partial differential relations satisfied by
these fields.  Solutions to the classical field theory are then
sections of the appropriate vector bundles that satisfy the partial
differential relations as well as regularity conditions (e.g.,
smoothness). Similarly, in our framework, a quantum field theory is
specified by providing the list of fields $\phi^{(i)}$ occuring in the
theory and the list of local, OPE relations satisfied by these fields.
Thus, the OPE relations play a role completely analogous to the role
of field equations in classical field theory. States---which are the
analogs of solutions in classical field theory---are positive linear
maps on the algebra $\A$ defined in section 3 that satisfy the OPE
relations as well as regularity conditions (in this case, the
microlocal spectrum condition). We note that in classical field
theory, the field equations always manifest all of the symmetries of
the theory, even in cases where there are no solutions that manifest
these symmetries.  Similarly, in our formulation of quantum field
theory, the OPE relations that define the theory should always respect
the symmetries of the theory~\cite{Weinberg}, even if no states
happen to respect these symmetries.

Our viewpoint on quantum field theory is more restrictive than
standard viewpoints in that we require the existence of an OPE. On the
other hand, it is less restrictive in that we do not require the
existence of a ground state. This latter point is best illustrated by
considering a free Klein-Gordon field $\varphi$ in Minkowski
spacetime
\ben
(\Box - m^2) \varphi = 0 \, ,
\een
where the mass term, $m^2$, is allowed to be positive,
zero, or negative. In the standard viewpoint, a quantum field theory
of the free Klein-Gordon field does
not exist in any dimension when $m^2 < 0$ and does not exist in
$D = 2$ when $m^2 = 0$ on account of the non-existence of a Poincare
invariant state. However, there is no difficulty is specifying OPE relations
that satisfy our axioms for all values of $m^2$ and all $D \geq 2$.
In particular, for $D=4$
we can choose the OPE-coefficient $C$ of the identity in the OPE of
$\varphi(x_1) \varphi(x_2)$ to be given by
\bena\label{kgid}
&&C(x_1, x_2; y) =\\
&&\frac{1}{4\pi^2} \left( \frac{1}{\Delta x^2 + i0t} +
m^2 \, j[m^2 \Delta x^2] \, \log [\mu^2 (\Delta x^2+i0t)] + m^2 \, h[m^2\Delta x^2] \right) \, , \non
\eena
where $\Delta x^2 = (x_1-x_2)^2$ and $t=x^0_1-x^0_2$.
Here $\mu$ is an arbitrarily chosen mass scale and
$j(z) \equiv \frac{1}{2i \sqrt{z}} J_1(i\sqrt{z})$ is an analytic function
of $z$, where $J_1$ denotes the Bessel function of order $1$.
Furthermore, $h(z)$ is the analytic function defined by
\ben
h(z) = -\pi \sum_{k=0}^\infty[\psi(k+1) + \psi(k+2)]\frac{(z/4)^k}{k!(k+1)!} \, .
\een
with $\psi$ the psi-function.
This formula for the OPE coefficient---as well as the corresponding
formulas for all of the other OPE coefficients---is as well defined for
negative $m^2$ as for positive $m^2$. Existence of states satisfying all
of the OPE relations for negative $m^2$ can be proven by the deformation
argument of~\cite{FNW}, using the fact that such states
exist for positive $m^2$.

Although, in our framework, the Klein-Gordon field with negative $m^2$
now joins the ranks of legitimate quantum field theories, this theory
is not physically viable because, in all states, field quantities will
grow exponentially in time\footnote{In this respect, the quantum field
of the Klein-Gordon field with negative $m^2$ behaves very similarly
to the corresponding classical theory. The classical Klein-Gordon
field with negative $m^2$ has a well posed initial value formulation
and causal propagation (despite frequently expressed claims to the
contrary). However, the classical Klein-Gordon field with negative
$m^2$ is not physically viable since it is unstable, i.e., it admits
solutions that grow exponentially with time.}. The potential
importance of the above example is that it explicitly demonstrates
that the local OPE coefficients can have a much more regular behavior
under variations of the parameters of the theory as compared with
state-dependent quantities, such as vacuum expectation values. The OPE
coefficients in the above example are analytic in $m^2$.  On the other
hand, the 2-point function of the global vacuum state is, of course,
defined only
for $m^2 \geq 0$ and is given by
\bena\label{kgvev}
&&\langle 0| \varphi(x_1) \varphi(x_2)|0 \rangle  =\\
&&\frac{1}{4\pi^2} \left( \frac{1}{\Delta x^2 +i0t} +
m^2 \, j[m^2 \Delta x^2] \, \log [m^2 (\Delta x^2+i0t)]
+ m^2 \, h[m^2\Delta x^2]
\right) \, . \non
\eena
This behaves non-analytically in $m^2$ at $m^2 = 0$ on account of the
$\log m^2$ term. In other words, in free Klein-Gordon theory, vacuum
expectation values cannot be constructed perturbatively by expanding
about $m^2 = 0$---as should be expected, since no vacuum state exists
for $m^2 < 0$---but there is no difficulty in perturbatively constructing
the OPE coefficients by expanding about $m^2 = 0$.

The above considerations raise the possibility that the well known
failure of convergence of perturbation series in interacting quantum
field theory may be due to the non-analytic dependence of states on
the parameters of the theory, rather than any non-analytic dependence
of the fields themselves, i.e., that the OPE coefficients may vary
analytically with the parameters of the theory. In other words, we are
suggesting the possibility that the perturbation series for OPE
coefficients may converge, and, thus, that, within our framework, it
may be possible to perturbatively construct\footnote{However, we are
not suggesting that it should be possible to
perturbatively construct states of the theory. Even if one had the complete
list of OPE coefficients, it would not be obvious how to construct states.}
interacting quantum field
theories. In order to do so, it will be necessary to define the basis
fields $\phi^{(i)}$ appropriately (see below and~\cite{HWnext})
and also to parametrize
the theory appropriately (since a theory with an analytic dependence
on a parameter could always be made to appear non-analytic by a
non-analytic reparametrization). Aside from the free Klein-Gordon
example above, the only evidence we have in favor of convergence of
perturbative expansions for OPE coefficients is the example of
super-renormalizable theories, such
as $\lambda \varphi^4$-theory in two spacetime
dimensions~\cite{hollandskopper}. Here, only finitely many terms in a
perturbative expansion can contribute to any OPE coefficient up to any
given scaling degree, so convergence (up to any given scaling degree)
is trivial. By contrast, for $\lambda \varphi^4$-theory in two
spacetime dimensions, the rigorously constructed, non-perturbative
ground state $n$-point functions can be proven to be non-analytic
at $\lambda=0$~\cite{Rivasseau}.

In cases---such as free Klein-Gordon theory above---where the OPE
coefficients {\it can} be chosen to be analytic in the parameters of
the theory, it seems natural to {\it require} that the theory be
defined so that this analytic dependence holds. This requirement has
some potentially major ramifications. Since vacuum expectation values of
a products of fields (i.e., a
correlation function) would be expected to have a
non-analytic dependence on the parameters of the theory, it follows that
if the OPE coefficients have an analytic dependence on these
parameters, then, even in Minkowski spacetime,
some of the fields appearing on the right side
of the OPE of a product
of fields must acquire a nonvanishing vacuum expectation value, at least for
some values of the parameters. This point is well illustrated by the above
Klein-Gordon example. It is natural to identify the next term
in the OPE of $\varphi(x_1) \varphi(x_2)$ [i.e., the
term beyond the identity term, whose coefficient is given by eq.~\eqref{kgid}]
as being $\varphi^2$ (with unit coefficient), i.e.,
\ben
\varphi(x_1) \varphi(x_2) \sim C(x_1, x_2; y) \myid
+ \varphi^2 (y) + ... \,\, .
\label{ps}
\een
This corresponds to the
usual ``point-splitting'' definition of $\varphi^2$, except that
$C(x_1, x_2; y)$ now replaces\footnote{The point-split expression
using $\langle 0|\varphi(x_1) \varphi(x_2) |0 \rangle$ yields the ``normal
ordered'' quantity $:\varphi^2:$. From the point of view of quantum field theory
in curved spacetime it is much more natural define $\varphi^2$
via eq.~\eqref{ps} than by normal ordering, since there is no generalization
of normal ordering to curved spacetime that is compatible with a local and
covariant definition of $\varphi^2$~\cite{hw1} Indeed, it follows from
the results of~\cite{hw1} that eq.~\eqref{ps} is the unique way to define
$\varphi^2$ compatible with desired properties, with the only ambiguities
in the definition of $\varphi^2$ arising from different allowed
choices of $C(x_1,x_2;y)$.}
$\langle 0|\varphi(x_1) \varphi(x_2) |0 \rangle$.
If we take
the vacuum expectation value of this formula (for $m^2 \geq 0$, when a
vacuum state exists)
and compare it with
eq.~\eqref{kgvev}, we obtain
\ben
\langle 0| \varphi^2(y) |0 \rangle = -\frac{m^2}{16\pi^2} \log (m^2/\mu^2) \, .
\een
Thus, we cannot set $\langle 0| \varphi^2 |0 \rangle = 0$ for all values
of $m^2$.
A similar calculation for the stress-energy tensor of $\varphi$ yields
\ben
\langle 0| T_{ab}(y) |0 \rangle = \frac{m^4}{64\pi^2} \log (m^2/\mu^2) \eta_{ab} \, .
\label{Tab}
\een

As in other approaches, the freedom to choose the arbitrary
mass scale $\mu$ in eq.~\eqref{kgid} gives rise to a freedom to
choose the value of the ``cosmological constant term''
in eq.~\eqref{Tab}.  However, unlike other
approaches, there is no freedom to adjust the value of the
cosmological constant when $m^2 = 0$ (i.e., we unambiguously obtain
$\langle 0| T_{ab} |0 \rangle = 0$ in Minkowski spacetime
in that case), and the $m^2$-dependence
of the cosmological constant is fixed (since $\mu$ is not allowed to
depend upon $m^2$).

A much more interesting possibility arises for interacting field
theories, such as non-abelian gauge theories. In such theories, it is
expected that there are ``non-perturbative effects'' that vary with
the coupling parameter $g$ as $\exp(-1/g^2)$.  Such non-perturbative
effects can potentially be very small compared with the natural scales
appearing in the theory. If such non-perturbative terms occur the
vacuum expectation values of products of fields and if---as we have
speculated above---the OPE coefficients have an analytic dependence on
the coupling parameter, then composite fields---such as the
stress-energy tensor---must acquire nonvanishing vacuum expectation
values that vary as $\exp(-1/g^2)$. This possibility appears worthy
of further investigation.

\bigskip

\noindent
{\bf Acknowledgments:}
This research was supported in part by NSF Grant PHY04-56619 to the University
of Chicago.

\appendix

\section{Definition of the scaling
degree and wave front set of a distribution}\label{appwfs}

In this appendix, we recall the notion of scaling degree and of the
wave front set of a distribution, which play an important role in
the body of the paper. Quite generally, let $u$ be an distribution on
$\mr^n$. We say that $u$ has scaling degree $d$ at $x=0$, if
$d$ is the smallest real number such that $\lambda^{\delta}
u(f_\lambda) \to 0$ as $\lambda \to 0+$, for all $\delta > d$. Here,
$f_\lambda(x) = \lambda^{-n} f(\lambda^{-1} x)$ denotes the function
of compact support that is obtained by rescaling a smooth test function $f$
around $x=0$, making it more and more sharply peaked at that point.
The scaling degree at an arbitrary point is obtained by simply
translating the distribution $u$ or the test function $f$ by the
desired amount. We write ${\rm sd}_x(u) = d$ for the scaling degree at
a point $x$.

We next recall the definition of the wave front set of a distribution
$u$ on $\mr^n$. Let $\chi$ be any smooth function of compact support.
Then $\chi u$ is evidently a distribution of compact support, and
its Fourier transform $\widehat{\chi u}(k)$ defines an entire function
of $k \in \mr^n$. For any distribution $v$ of compact support, we
define its corresponding ``singular set'', $\Sigma(v)$ as the
collection of all $k \in \mr^n$ such that
\ben
|\widehat v(\lambda k)| \ge C \lambda^N \, ,
\een
for some $C>0$, and some $N$, and all $\lambda >0$. We define the wave
front set $\WF_x(u)$ at a point $x \in \mr^n$ as the intersection
\ben
\WF_x(u) = \bigcap_{\chi: x \in \supp \chi} \Sigma(\chi u) \, ,
\een
and we define $\WF(u)$ as the union
\ben
\WF(u) = \bigsqcup_{x \in \mr^n} \WF_x(u) \, .
\een
Each set $\WF_x(u)$ is a conic set, in the sense that if $k \in
\WF_x(u)$, then so is $tk$ for any $t>0$, and $k=0$ is never in
$\WF_x(u)$. It immediately follows from the definition that
$\WF_x(u) = \emptyset$ if and only if $u$ can be represented by
a smooth function in an open neighborhood of $x$. In this sense,
the wave-front set tells one at which points a distribution is
singular. It also contains information about the most singular
directions in local momentum space, which are represented by
$k \in \WF_x(u)$.

It turns out that both the scaling degree of a distribution at a point
$x$, as well as the wave front set at $x$ are invariantly defined. By
this one means the following. Let $\rho: V \to U$ be a smooth
diffeomorphism between open sets $U,V \subset \mr^n$. Let
$u$ be a distribution supported in $U$, and let $\rho^* u$ be the
pulled back distribution in $V$, where the pull-back is defined
by analogy with the pull back of a smooth function. Then it is easy to
show that $\sd_x(\rho^* u) = \sd_{\rho(x)}(u)$. Furthermore, if
$x' = \rho(x)$, and if we define $\rho^*(x',k') = (x, k)$, where
$k=[d\rho(x)]^{\rm v} k'$, then one can show
\ben
\WF_x(\rho^* u) = \rho^* \WF_{\rho(x)}(u) \, .
\een
These relations imply that the scaling degree and the wave front set
can be invariantly defined on an arbitrary manifold $X$,
and that the wave-front set should be viewed as a subset of
$T^* X$.

In the body of the paper, we frequently consider the
case $X=M^{n+1}$, and the scaling degree at the point $(y,y,\dots,
y)$, i.e., points on the total diagonal. To save writing, this is
simply denoted $\sd \, u$.

\section{Equivalent formulation of condition (C6)}

In this appendix, we relate the scaling degree (C5) and
asymptotic positive (C6) assumptions
to other properties of the quantum field theory. Our
first is just a repetition of a result obtained in~\cite{Fewster}:

\begin{thm} For any $x \in M$, and any (scalar) field $T$ not equal
to a multiple of the identity operator,
we define the convex set $S_x \subset \mr$ by
\ben
S_x = \{\langle T(x)\rangle_\Phi \mid \text{$\Phi$ a normalized state}\}
\een
Then, $S_x = \mr$ for at least one spacetime $\M$.
\end{thm}

\medskip
\noindent
\paragraph{Remark:} The statement means that pointlike hermitian fields $T(x)$ are
unbounded from above {\em and} below, even though their classical
counterpart (if the theory has a classical limit) might be manifestly
non-negative, such as the Wick square $T=\varphi^2$, or the energy density
$T = T_{ab} u^a u^b$ of a free Klein-Gordon field
$\varphi$.

\medskip
\noindent
{\em Proof}:
Choose any state $\Phi$. We may
assume that $\langle T(f) \rangle_\Phi = 0$, because if not, we just
need to consider instead $T(x)$ by $T(x) - \langle T(x) \rangle_\Phi
\myid$. Define
\ben
A = \cos \alpha \, \myid + \sin \alpha \,
\frac{T(f)}{\langle T(f)T(f) \rangle_\Phi^{1/2}}
\, ,
\een
and define a new normalized state by $\langle \, . \rangle_\alpha =
\langle A \, . \,
A^* \rangle_\Phi$. Then
\ben
\langle T(f) \rangle_{\alpha} = a \sin 2\alpha + b(1-\cos 2\alpha)
\,
\een
where
\ben
a = \langle T(f)T(f) \rangle_\Phi^{1/2} \, , \quad b =
\frac{1}{2} \frac{\langle T(f)T(f)T(f) \rangle_\Phi}{
\langle T(f)T(f)\rangle_\Phi} \, .
\een
Minimizing over $\alpha$ gives
\ben
\inf_{\Psi} \, \langle T(f) \rangle_\Psi \le b - \sqrt{a^2 + b^2} \, .
\een
Replacing $f$ by $-f$ also gives
\ben
\sup_{\Psi} \, \langle T(f) \rangle_\Psi \ge -b + \sqrt{a^2 + b^2} \, .
\een
Now, consider a test function $f$ with $\int f d\mu = 1, f(x) \neq 0$.
By lemma~\ref{scalelemma}, there exists a spacetime and a state such
that $\langle T(\bar f_\lambda) T(f_\lambda) \rangle_\Phi \to +\infty$ for
some appropriately chosen $f$, where $f_\lambda(y) =
\lambda^{-D} f(x + \lambda^{-1}(y-x))$. Hence $a \to \infty$ as $\lambda \to 0$.
and therefore $\inf_\Psi \langle T(f_\lambda) \rangle_\Psi$ becomes arbitrarily
small as $\lambda \to 0$. On the other hand $f_\lambda \to \delta_x$ in this
limit, so the set $S_x$ is not bounded below. It similarly follows that
it cannot be bounded above either. \qed

\medskip
\noindent
Our next result is in some sense a converse to the above result:

\begin{thm}
Let the set $S_x$ be equal to $\mr$ for a given spacetime $\M$, and
all hermitian operators $T$ not equal to a multiple of the identity.
Then for any $i,k \in I$ with $i \neq \myid$, and any sections $v_{(i)}$ of
$V(i)$ we have that
\ben\label{scaleineq}
\sd [C^{(k)(k^\star)}_{(i)} (v_{(k)} \otimes \bar v_{(k^\star)})]
<
\sd [C^{(k)(k^\star)}_{(\myid)} (v_{(k)} \otimes \bar v_{(k^\star)})] \, .
\een
\end{thm}

\medskip
\noindent
\paragraph{Remark:} In generic spacetimes, we expect that the scaling degree
of the right side is equal to $2\d(k)$, where $\d(k)$ is the dimension of
the field $\phi^{(k)}$; see the scaling degree axiom (C5). We also expect the
quantity on the left side to be equal to $2\d(k) - \d(i)$; see again (C5).
Thus, the result tells us that, in this situation, $\d(i) > 0$ unless
$\phi^{(i)}$ is the identity field. Thus, in this sense, the assumption of the
theorem implies the
asymptotic positivity axiom (C6), or---stated differently---the
asymptotic positivity axiom is inconsistent with not having $S_x = \mr$.

\medskip
\noindent
{\em Proof}:
By assumption, we can find
a state $\Phi$ such that
$\langle T(x) \rangle_{\Phi} > A$ for each $A \in \mr$.
Consider an arbitrary, but fixed,
finite collection $\phi^{(1)}, \dots, \phi^{(n)}$
of fields. Each field is valued in some vector bundle $V(i)$. The
set of expectation values of this collection of fields forms a subset
which we denote
\ben
K_x = \bigg\{
( \langle \phi^{(1)}(x) \rangle_\Phi, \dots, \langle \phi^{(n)}(x) \rangle_\Phi )
\bigg| \,\,\, \text{states $\Phi$} \bigg\}
\subset \bigoplus_{i=1}^n V(i)_x  =: V_x \, .
\een
Because the set of states is convex (i.e., any convex linear
combination of normalized states is again a normalized state), the set
$K_x$ is a convex subset of $V_x$.
We claim that, in fact, $K_x = V_x$. Assume that this were
not the case. Then, since any {\em proper} convex subset of a finite
dimensional vector space can be enclosed by a collection of planes,
there exists a collection of dual vectors $c_{(i)} \in V(i)^{\rm v}$
and an $A \in \mr$ such that
\ben
{\rm Re}\sum_{(i)} c_{(i)} v^{(i)} < A \, , \quad \text{for all $(v^{(1)},
  \dots, v^{(n)}) \in K_x$.}
\een
However, this would mean by definition that if
$T = {\rm Re} \sum c_{(i)} \phi^{(i)}$, then $\langle T(x)
\rangle_\Phi < A$ for all states $\Phi$, a contradiction.

Assume that the statement of the theorem is not true. Let $i=1, \dots, n \in I$ be
the field labels, with $i \neq \myid$
for which the inequality~\eqref{scaleineq}
does not hold. From what we have just shown, if we define $K_x$ as above,
then $K_x$ is equal to $V_x$. In particular, we may find states
$\Phi, \Psi$, and nonzero $v^{(i)}$ with the property that
\ben
\langle \phi^{(i)}(x) \rangle_\Psi = v^{(i)}, \quad
\langle \phi^{(i)}(x) \rangle_\Phi = -v^{(i)} \, .
\een
Let $f_\lambda(x) = \lambda^{-D} f(y+\lambda^{-1}(x-y))$ be a test section
in the dual of the space $V(k)$, and let $\delta$ be a real number
which is bigger than $2\d(k)$, but smaller than the left side
of eq.~\eqref{scaleineq}. Using the fact that $\langle \phi^{(k)} (f_\lambda)
\phi^{(k^\star)}(\bar f_\lambda) \rangle_\Psi \ge 0$ for all $\lambda$,
we find from the operator product expansion that
\ben
\lambda^{\delta} \sum_{(i)} C^{(k^\star)(k)}_{(i)}(\bar f_\lambda,
f_\lambda; x) v^{(i)}_{} \to + \infty \, ,
\een
for at least one $f$ and a subsequence of $\lambda$ tending to $0$.
Applying a similar argument to the state $\Phi$ gives that
\ben
-\lambda^{\delta}
\sum_{(i)} C^{(k^\star)(k)}_{(i)}(\bar f_\lambda, f_\lambda; x) v^{(i)}_{}
\to +\infty \, ,
\een
for this subsequence of $\lambda$ tending to $0$.
This is a contradiction, so the inequality~\eqref{scaleineq} must hold.
\qed


\begin{thebibliography}{99}

\bibitem{Singer1990}
  S.~Axelrod and I.~M.~Singer,
  ``Chern-Simons perturbation theory. 2,''
  J.\ Diff.\ Geom.\  {\bf 39}, 173 (1994)

\bibitem{vertex1}
 R.~E.~Borcherds,
  ``Vertex Algebras, Kac-Moody Algebras, And The Monster,''
  Proc.\ Nat.\ Acad.\ Sci.\  {\bf 83}, 3068 (1986).

\bibitem{Bostelmann1}
  H.~Bostelmann,
  ``Operator product expansions as a consequence of phase space properties,''
  J.\ Math.\ Phys.\  {\bf 46}, 082304 (2005);

\bibitem{Bostelmann2}
H. Bostelmann: ``Phase space properties and the short distance structure in quantum  field
  theory,''
  J.\ Math.\ Phys.\  {\bf 46}, 052301 (2005)

\bibitem{bfk} R. Brunetti, K. Fredenhagen and M. K\"ohler: ``The microlocal spectrum
condition and Wick polynomials on curved spacetimes,'' Commun. Math.
Phys. {\bf 180}, 633-652 (1996)

\bibitem{bf} R. Brunetti and K. Fredenhagen: ``Microlocal Analysis and
Interacting Quantum Field Theories:
Renormalization on physical backgrounds,''
Commun. Math. Phys. {\bf 208}, 623-661 (2000)

\bibitem{bfv}
R.~Brunetti, K.~Fredenhagen and R.~Verch,
``The generally covariant locality principle: A new paradigm for local  quantum physics,''
Commun.\ Math.\ Phys.\  {\bf 237}, 31 (2003),
[math-ph/0112041]; see also K.~Fredenhagen,
``Locally covariant quantum field theory,'' [arXiv:hep-th/0403007].

\bibitem{Fewster}
C. J. Fewster: "Energy inequalities in quantum field theory,"
Proceedings of XIVth International Congress on Mathematical Physics, ed. J.-C. Zambrini,
559 (2003)

\bibitem{Fredlisbon}
K. Fredenhagen: "Locally covariant quantum field theory,"
Proceedings of XIVth International Congress on Mathematical Physics, ed. J.-C. Zambrini,
29 (2003)

\bibitem{fre}
 K.~Fredenhagen and J.~Hertel,
  ``Local Algebras Of Observables And Point - Like Localized Fields,''
  Commun.\ Math.\ Phys.\  {\bf 80}, 555 (1981);
  K.~Fredenhagen and M.~Jorss,
  ``Conformal Haag-Kastler nets, point - like localized fields and the
  existence of operator product expansions,''
  Commun.\ Math.\ Phys.\  {\bf 176}, 541 (1996).

\bibitem{vertex3}
  I.~Frenkel, J.~Lepowsky and A.~Meurman,
  ``Vertex operator algebras and the Monster,''
Academic Press, Boston (1988)

\bibitem{FNW}
S.A. Fulling, F.J. Narcowich, and R.M. Wald: ``Singularity Structure
of the Two-Point Function in Quantum Field Theory in Curved Spacetime, II,''
Ann. Phys. {\bf 136} 243 (1981).

\bibitem{Fultonmac}
W. Fulton and R. MacPherson: ``A compactification of configuration
spaces,'' Ann. Math. {\rm 139} 183 (1994)

\bibitem{haag}
  R.~Haag and D.~Kastler,
  ``An Algebraic Approach To Quantum Field Theory,''
  J.\ Math.\ Phys.\  {\bf 5}, 848 (1964).


\bibitem{hollands2006}
  S.~Hollands,
  ``The operator product expansion for perturbative quantum field theory in
  curved spacetime,''
  Commun.\ Math.\ Phys.\  {\bf 273}, 1 (2007)
  [arXiv:gr-qc/0605072].


\bibitem{hpct}
  S.~Hollands,
  ``A general PCT theorem for the operator product expansion in curved
  spacetime,''
  Commun.\ Math.\ Phys.\  {\bf 244}, 209 (2004)
  [arXiv:gr-qc/0212028].

\bibitem{hollandskopper}
S. Hollands and C. Kopper: in progress


\bibitem{HWnext}
S. Hollands and R.M. Wald: in progress

\bibitem{hw1} S. Hollands and R. M. Wald: ``Local Wick Polynomials and Time Ordered Products
of Quantum Fields in Curved Space,'' Commun. Math. Phys. {\bf 223}, 289-326 (2001), [gr-qc/0103074]

\bibitem{hw2}
S.~Hollands and R.~M.~Wald:
``Existence of local covariant time-ordered-products of quantum fields in  curved spacetime,'' Commun. Math. Phys. {\bf 231}, 309-345 (2002), [gr-qc/0111108]

\bibitem{hollands08}
 S.~Hollands,
  ``Quantum field theory in terms of consistency conditions I: General
  framework, and perturbation theory via Hochschild cohomology,''
  arXiv:0802.2198 [hep-th].

\bibitem{hormander}
L. H\"ormander: {\it The Analysis of Linear Partial Differential Operators I,} Berlin, Springer-Verlag 1983

\bibitem{vertex2}
  V.~Kac,
  ``Vertex algebras for beginners,''
{ Providence, USA: AMS (1996) 141 p. (University lectures series. 10)}


\bibitem{radzikowski}
  M.~J.~Radzikowski,
  ``Micro-local approach to the Hadamard condition in quantum field theory on
  curved space-time,''
  Commun.\ Math.\ Phys.\  {\bf 179}, 529 (1996).

\bibitem{Rivasseau}
  V.~Rivasseau,
  ``From perturbative to constructive renormalization,''
Princeton, USA: Univ. Pr. (1991) 336 p. (Princeton series in physics)



\bibitem{schroer}
  B.~Schroer, J.~A.~Swieca and A.~H.~Volkel,
  ``Global Operator Expansions In Conformally Invariant Relativistic Quantum
  Field Theory,''
  Phys.\ Rev.\ D {\bf 11}, 1509 (1975).

\bibitem{sw} R. F. Streater and A. A. Wightman: {\it PCT, Spin and Statistics and All That,} New York, Benjamin 1964


\bibitem{wald} R. M. Wald: {\it Quantum Field Theory on Curved Spacetimes and Black Hole Thermodynamics,}
The University of Chicago Press, Chicago (1990)

\bibitem{wilson}
 K.~G.~Wilson,
  ``Nonlagrangian Models Of Current Algebra,''
  Phys.\ Rev.\  {\bf 179}, 1499 (1969).

\bibitem{Weinberg}
See appendix in: C. Bernard, A. Duncan, J. LoSecco, S. Weinberg: ``Exact spectral-function sum rules'',
Phys. Rev. {\bf D 12}, 792 - 804 (1975)

\bibitem{zimmermann}
  W.~Zimmermann,
  ``Normal Products And The Short Distance Expansion In The Perturbation Theory
  Of Renormalizable Interactions,''
  Annals Phys.\  {\bf 77}, 570 (1973)
  [Lect.\ Notes Phys.\  {\bf 558}, 278 (2000)].

\end{thebibliography}
\end{document}